\PassOptionsToPackage{prologue,dvipsnames}{xcolor}
\documentclass[conference]{IEEEtran}

\usepackage{algorithmicx,algorithm}
\usepackage{amsmath}
\usepackage{amssymb}
\usepackage{amsthm}
\usepackage{array}
\usepackage[noend]{algpseudocode}
\usepackage{balance}
\usepackage{cite}
\usepackage[dvipsnames]{xcolor}
\usepackage{enumerate}
\usepackage{etoolbox,paralist}
\usepackage{float}
\usepackage{graphicx}
\usepackage{hyperref}
\usepackage{cleveref}
\usepackage{listings}
\usepackage{makecell}
\usepackage{multirow}
\usepackage{pifont}
\usepackage{setspace}
\usepackage{subcaption}
\usepackage{textcomp}
\usepackage{threeparttable}
\usepackage{url}
\usepackage{tikz,balance}

\hypersetup{
    colorlinks=true,
    linkcolor=ForestGreen,
    citecolor=blue,
    filecolor=magenta,      
    urlcolor=.,
    pdftitle={Overleaf Example},
    pdfpagemode=FullScreen,
    }

\makeatletter
\patchcmd{\@makecaption}
  {\scshape}
  {}
  {}
  {}
\makeatother

\definecolor{verylightgray}{rgb}{.97,.97,.97}

\lstdefinelanguage{Solidity}{
	keywords=[1]{anonymous, assembly, assert, balance, break, call, callcode, case, catch, class, constant, continue, constructor, contract, debugger, default, delegatecall, delete, do, else, emit, event, experimental, export, external, false, finally, for, function, gas, if, implements, import, in, indexed, instanceof, interface, internal, is, length, library, log0, log1, log2, log3, log4, memory, modifier, new, payable, pragma, private, protected, public, pure, push, require, return, returns, revert, selfdestruct, send, solidity, storage, struct, suicide, super, switch, then, this, throw, transfer, true, try, typeof, using, value, view, while, with, addmod, ecrecover, keccak256, mulmod, ripemd160, sha256, sha3}, 
	keywordstyle=[1]\color{blue}\bfseries,
	keywords=[2]{address, bool, byte, bytes, bytes1, bytes2, bytes3, bytes4, bytes5, bytes6, bytes7, bytes8, bytes9, bytes10, bytes11, bytes12, bytes13, bytes14, bytes15, bytes16, bytes17, bytes18, bytes19, bytes20, bytes21, bytes22, bytes23, bytes24, bytes25, bytes26, bytes27, bytes28, bytes29, bytes30, bytes31, bytes32, enum, int, int8, int16, int24, int32, int40, int48, int56, int64, int72, int80, int88, int96, int104, int112, int120, int128, int136, int144, int152, int160, int168, int176, int184, int192, int200, int208, int216, int224, int232, int240, int248, int256, mapping, string, uint, uint8, uint16, uint24, uint32, uint40, uint48, uint56, uint64, uint72, uint80, uint88, uint96, uint104, uint112, uint120, uint128, uint136, uint144, uint152, uint160, uint168, uint176, uint184, uint192, uint200, uint208, uint216, uint224, uint232, uint240, uint248, uint256, var, void, ether, finney, szabo, wei, days, hours, minutes, seconds, weeks, years},	
	keywordstyle=[2]\color{teal}\bfseries,
	keywords=[3]{block, blockhash, coinbase, difficulty, gaslimit, number, timestamp, msg, data, gas, sender, sig, value, now, tx, gasprice, origin},	
	keywordstyle=[3]\color{violet}\bfseries,
	identifierstyle=\color{black},
	sensitive=false,
	comment=[l]{//},
	morecomment=[s]{/*}{*/},
	commentstyle=\color{violet}\ttfamily,
	stringstyle=\color{red}\ttfamily,
	morestring=[b]',
	morestring=[b]"
}

\makeatletter
\newcommand{\removelatexerror}{\let\@latex@error\@gobble}

\newtheorem{definition}{Definition}
\newtheorem{theorem}{Theorem}
\newtheorem{lemma}{Lemma}
\algnewcommand\algorithmicforeach{\textbf{for each}}
\algdef{S}[FOR]{ForEach}[1]{\algorithmicforeach\ #1\ \algorithmicdo}

\newenvironment{myproof}[1]{\noindent \emph{Proof{#1}:}}{\hfill$\square$}

\algnewcommand\algorithmicinput{\textbf{INPUT:}}
\algnewcommand\INPUT{\item[\algorithmicinput]}
\algnewcommand\algorithmicoutput{\textbf{OUTPUT:}}
\algnewcommand\OUTPUT{\item[\algorithmicoutput]}
\algnewcommand\algorithmicassume{\textbf{ASSUME:}}
\algnewcommand\ASSUME{\item[\algorithmicassume]}

\begin{document}


\title{Optimal Sharding for Scalable Blockchains with Deconstructed SMR}





\author{
\IEEEauthorblockN{Jianting Zhang\textsuperscript{$*$}, Zhongtang Luo\textsuperscript{$*$}, Raghavendra Ramesh\textsuperscript{$\dag$}, and Aniket Kate\textsuperscript{$*$$\dag$}}
\IEEEauthorblockA{\textsuperscript{$*$}Purdue University, \textsuperscript{$\dag$}Supra Research \\
}
\\
}

\IEEEoverridecommandlockouts
\makeatletter\def\@IEEEpubidpullup{6.5\baselineskip}\makeatother

\maketitle



\begin{abstract} 
Sharding enhances blockchain scalability by dividing system nodes into multiple shards to handle transactions in parallel. However, a size-security dilemma where every shard must be large enough to ensure its security constrains the efficacy of individual shards and the degree of sharding itself. Most existing sharding solutions therefore rely on either weakening the adversary or making stronger assumptions on network links.

This paper presents Arete, an optimally scalable blockchain sharding protocol designed to resolve the dilemma based on an observation that if individual shards can tolerate a higher fraction of (Byzantine) faults, we can securely create smaller shards in a larger quantity. The key idea of Arete, therefore, is to improve the security resilience/threshold of shards by dividing the blockchain's State Machine Replication (SMR) process itself. Similar to modern blockchains, Arete first decouples SMR in three steps: transaction dissemination, ordering, and execution. However, unlike other blockchains, for Arete, a single ordering shard performs the ordering task while multiple processing shards perform the dissemination and execution of blocks. As processing shards do not run consensus, each of those can tolerate up to half compromised nodes. Moreover, the SMR process in the ordering shard is extremely lightweight as it only operates on the block digests. Second, Arete considers safety and liveness against Byzantine failures separately to improve the safety threshold further while tolerating temporary liveness violations in a controlled manner. Apart from the creation of more optimal-size shards, such a deconstructed SMR scheme also empowers us to devise a novel certify-order-execute architecture to fully parallelize transaction handling, thereby significantly improving the performance of sharded blockchain systems. We implement Arete and evaluate it on a geo-distributed AWS environment by running up to 500 nodes. Our results demonstrate that Arete outperforms the state-of-the-art sharding protocols in terms of scalability, transaction throughput, and cross-shard confirmation latency without compromising on intra-shard confirmation latency.

\end{abstract}

\section{Introduction}\label{section-introduction}

Scalability is one of the major issues facing blockchain technology today. 
Sharding is proposed to address the scalability issue~\cite{Elastico, Omniledger, Rapidchain, gearbox, ahl, sharper, byshard, zhang2023front}.
By dividing nodes into multiple groups (i.e., \emph{shards}) and enabling distinct shards to handle transactions with an \emph{independent, intra-shard} consensus or state machine replication (SMR) process, blockchain sharding can achieve an increased transaction throughput as more nodes join the system. Due to the decreasing size of the consensus group, one primary concern of sharding is how to form secure shards, where the ratio of Byzantine nodes of a shard cannot exceed the security threshold $f$ that an intra-shard consensus can tolerate. 
For instance, it is well-known that the classical non-synchronous Byzantine Fault-Tolerant (BFT) consensus/SMR can only tolerate up to $1/3$ corruptions, i.e., $f<1/3$~\cite{DLT}. 

\noindent \textbf{The Size-security dilemma}. 
Unfortunately, the sharding protocol faces a size-security dilemma that constrains the shard size and the overall number of shards the system can create.
To elaborate, in blockchain sharding, having smaller shards means more shards~\cite{Rapidchain, gearbox, licochain, instachain}.
When dividing nodes into shards, a small shard size enables more shards to be created, but increases the probability of forming an insecure shard with higher than $f$ fraction of malicious nodes. Larger shards are more secure but only allow fewer shards to be created. The trade-off between shard size and the probability of forming secure enough shards is fundamental to securely scaling the blockchain systems via sharding. For instance, to achieve a $30$-bit statistical security\footnote{A 30-bit security level here means that a shard-forming probability with a Byzantine node ratio exceeding the security threshold is $\leq2^{-30}$.} in a system with $n=1000$ nodes and up to $s=1/4$ fraction of Byzantine nodes, the sharding system must sample $486$ nodes to form a shard with $f<1/3$ security threshold (Table~\ref{table:related work}). As only two shards can be created with the shard size $m=486$ and $n=1000$, the system scales poorly. 
Moreover, a large shard suffers from performance degradation when running most consensus protocols due to quadratic communication overheads~\cite{CivitDGGGKV24}. 
Therefore, an exciting and promising research direction in blockchain sharding is to securely reduce the shard size so that more and smaller shards can be created to enhance the efficacy and scalability of sharding.


\noindent \textbf{Existing solutions and limitations}.
To reduce the shard size while ensuring the security of the system, many existing solutions make stronger assumptions~\cite{Omniledger, Rapidchain, ahl, hong2021pyramid}. For instance, RapidChain~\cite{Rapidchain} assumes the network to be bounded synchronous, adopts synchronous consensus protocols~\cite{chan2020streamlet, abraham2020sync}, and enhances the security threshold for every shard to $f<1/2$. An enhanced security threshold enables shards to be smaller.
However, for any low latency and high-throughput blockchain system over the Internet, the bounder-synchronous assumption can be unrealistic. Some recent works~\cite{instachain, licochain, gearbox} advocate reducing the shard size by allowing more than $f$ Byzantine nodes to be sampled into a shard during the shard formation stage and resolve/recover it once the security-violated shard is detected. However, the recovery process is costly, and if security violations happen frequently, the sharding system can be inactive for a long duration while recovering security-violated shards. For instance, when \textsc{GearBox}~\cite{gearbox} can achieve the shard size $m=72$, the probability of a shard violating liveness reaches up to $65.78\%$. \textsc{GearBox} recovers a liveness-violated shard by injecting more nodes into the shard, but this recovery process takes several hours (or even a couple of weeks~\cite{ethereum-sync}), delaying shard availability until nodes synchronize the states. Such frequent security violations significantly diminish the system's appeal in practice.
\subsection{Arete Overview and Contributions}
This work presents Arete,
a highly scalable sharding protocol allowing the creation of lightweight shards securely. The key idea is to \textit{reduce the shard size by securely increasing the security threshold $f$} without making any new assumptions or reducing the adversarial capacity. To this end, Arete consists of two building blocks.

\noindent\textbf{(1) Ordering-processing sharding scheme.} 
Previous sharding protocols employ every shard to process (i.e., disperse and execute) and order transactions. Under partial synchrony/asynchrony, this upper-bounds the security threshold $f$ of shards to $1/3$ as every shard needs to run consensus. We observe that, while every shard should perform its data dissemination and execution, there is no need to ask every shard to order its transactions by itself; the ordering task though requires a supermajority of honest nodes as compared to transaction processing tasks, it can be extremely lightweight by replacing transactions/blocks by signed-hash digests. 

Specifically, the blockchain process can be divided into three repeatedly performed tasks: data dissemination/availability, ordering, and execution. The data availability task realizes that every honest node can reconstruct an intact block if the block is finalized. The ordering task realizes that all honest nodes must output the same transaction order. The execution task realizes that the SMR has the correct output state after executing the ordered transactions. Assuming the network is not synchronous, while the ordering task requires $f<1/3$ to tolerate equivocations performed by Byzantine nodes (e.g., voting for two conflicting blocks), the data availability and execution tasks allow $f<1/2$ as there is no equivocation problem for them. Moreover, the ordering task only requires slight resources in communication and computation (such as broadcasting and handling metadata of a block), but the data availability and execution tasks are always resource-intensive. For instance, the data availability and execution tasks require extensive bandwidth, computation, and storage to disperse and execute intact transactions, which have also been shown to be the bottleneck of SMR in recent works~\cite{keidar2021all, danezis2022narwhal, spiegelman2022bullshark, shoal}. 

Based on the above key insights, we propose a novel ordering-processing sharding architecture to increase the security threshold $f$ of shards, allowing smaller shard sizes. The ordering-processing sharding scheme completely decouples data dissemination, ordering, and execution of SMR, and shards these tasks based on their required security thresholds and resources. To elaborate, there are two types of shards in Arete: a single \emph{ordering shard} and multiple \emph{processing shards}. The ordering shard runs a BFT consensus protocol to determine a global transaction order, tolerating up to $1/3$ Byzantine nodes. Processing shards are responsible for dispersing intact transactions and executing the globally ordered transactions, each tolerating up to $1/2$ Byzantine nodes. A higher threshold enables us to form more processing shards with smaller sizes, eventually enhancing the scalability of our shared system. Beneficial from the ordering-processing sharding scheme, Arete adopts a \textit{certify-order-execute (COE)} architecture to efficiently finalize transactions (\S~\ref{sec-coe-model}). The COE architecture consists of three stages corresponding to the three tasks in the SMR. Briefly, each processing shard first certifies transactions by ensuring at least one honest node has intact transaction data (performing data availability task). The ordering shard then runs consensus to establish a global order for these certified transactions (performing ordering task). Finally, processing shards execute and finalize these globally ordered transactions (performing execution task). 
It is worth noting that we are not the first to explore decoupling technology for blockchains, and many previous works (partially) decouple SMR to achieve high performance. However, they either fail to be scalable to accommodate (even) hundreds of nodes, or have to make stronger trust assumptions. Arete is the first decoupling architecture that is highly scalable with no extra trust assumptions. We provide a detailed comparison between Arete and related works in \S~\ref{main-body-related-work}.
Notably, our ordering-processing sharding and the COE architecture bring five useful features:
\begin{compactitem}[-]
    \item \textbf{F1: Improved fault tolerance}. Without running a consensus, processing shards tolerate up to half Byzantine nodes. A higher threshold allows the creation of more lightweight shards. For instance, with $n=1000$, Arete can create $13$ shards while Omniledger~\cite{Omniledger} can only create two shards.
    \item \textbf{F2: Asynchronous deterministic processing}. Without a consensus task, processing shards operate completely asynchronously without making any timing assumptions and requiring randomization to avoid the FLP impossibility~\cite{fischer1985impossibility}.
    \item \textbf{F3: Lightweight ordering}. Nodes in the ordering shard neither receive transactions nor compute states for transactions. When only needed to handle lightweight signed digest, the ordering shard is scalable to service more processing shards. Our theoretical analysis (\S~\ref{sec-coe-model-optimization}) shows the system needs \textit{at least} $283$ processing shards before the ordering shard handles data volume similar to a processing shard. 
    \item \textbf{F4: Shard autonomy}. Processing shards disseminate and execute transactions independently without interfering with each other. Each of them can autonomously utilize their resources as the ordering shard orders varying numbers of transactions for them based on their speeds in processing transactions. 
    \item \textbf{F5: Lock-free execution}. Cross-shard transactions are notorious in sharding systems as they involve data managed by multiple shards and introduce a cost cross-shard protocol (e.g., the lock-based two-phase commit). In Arete, with a global order in the ordering shard, processing shards can consistently execute cross-shard transactions without locking relevant states. This significantly enhances the practicality of Arete, particularly in scenarios where popular smart contracts (e.g., cryptocurrency exchanges~\cite{coinbase} and Uniswap~\cite{uniswap}) are frequently accessed.
\end{compactitem}

\noindent\textbf{(2) Safety-liveness separation.}
Next, motivated by \textsc{GearBox}~\cite{gearbox}, Arete separates the safety and liveness of processing shards to further increase the safety threshold. 
Informally, safety indicates any two honest nodes store the same prefix of a transaction ledger, and liveness indicates the shard is available to handle transactions. 
In Arete, the security of a processing shard is defined as a tuple $(f_S, f_L)$, in which: (i) $f_S$ represents the safety threshold, ensuring safety as long as no more than $f_S$ Byzantine nodes exist in the shard; (ii) $f_L$ represents the liveness threshold, ensuring liveness as long as no more than $f_L$ Byzantine nodes exist in the shard. The safety-liveness separation empowers us to adjust $f_S$ and $f_L$ to offer a higher $f_S$ fraction. Specifically, the security thresholds in processing shards are set to $f_S > f_L$. When forming processing shards, Arete allows no more than $f_S$ (but can exceed $f_L$) Byzantine nodes in a processing shard, which can reduce the size of processing shards due to a larger $f_S$. In this case, a processing shard ensures safety statistically but could temporarily violate liveness. The temporary existence of liveness violation is admissible because, in a real-world system, 
temporary unavailability can be caused by benign reasons, such as routine maintenance and upgrades. 
Remarkably, Arete can still guarantee the liveness of the system eventually even though liveness-violated processing shards are formed.
This is because our ordering shard can detect and recover liveness-violated processing shards, making all processing shards eventually available to handle transactions.

The safety-liveness separation, however, inevitably forms liveness-violated shards, and the system needs to recover these shards. Fortunately, we notice that recovering shards would become less costly if it happens rarely. To evaluate the cost of shard recovery, we define a new property, called \textit{$\mathcal{P}$-probabilistic liveness} (\S~\ref{subsection-defnitions}), for the sharding system that separates safety and liveness (e.g., \textsc{GearBox}~\cite{gearbox}). Informally, it represents the probability of creating a new shard that guarantees liveness. In a practical scenario, a high $\mathcal{P}$-probabilistic liveness is acceptable. For instance, Amazon Web Service claims that $0.9999$-probabilistic availability (i.e., $0.9999$-probabilistic liveness as we claim in \S~\ref{appendix-analysis-p-probability}) is deemed acceptable for a commercial corporation~\cite{fournine}. Therefore, Arete significantly reduces the cost of shard recovery caused by the safety-liveness separation. Notably, since processing shards can tolerate $f<1/2$ corruptions, with safety-liveness separation, Arete only requires $f_S + f_L < 1$ under a partially synchronous/asynchronous model (see \S~\ref{section-preliminaries-sl-separation} for more details), whereas the safety-liveness separation used in \textsc{GearBox} necessitates $f_S + 2f_L < 1$. As shown in Table~\ref{table:related work}, although both Arete and \textsc{GearBox} can achieve the shard size $m=72$ by adjusting $f_S=0.57$, Arete ensures $0.9999$-probabilistic liveness while \textsc{GearBox} only achieves $0.3422$-probabilistic liveness.

In summary, we make the following contributions:
\begin{asparaitem}
        \item We propose Arete, a highly scalable blockchain sharding protocol allowing the system to create optimal-size shards to horizontally scale transaction handling (\S~\ref{section-overview}). Compared to existing solutions, Arete enhances the blockchain scalability without weakening the resilience to Byzantine failure or making synchrony assumptions, while maintaining low shard recovery costs.
    \item We propose a new COE architecture to efficiently finalize transactions (\S~\ref{sec-coe-model}). Empowered by our sharding protocol, the COE architecture can fully harness the resources of a large number of nodes, bringing remarkable performance enhancements in terms of transaction throughput and cross-shard confirmation latency.
    \item We provide a full implementation of Arete and compare it with two representative sharding protocols: \textsc{GearBox}~\cite{gearbox} that separates safety and liveness and RIVET~\cite{rivet} that separates the ordering shard from processing shards. We conducted experiments under a geo-distributed AWS environment (\S~\ref{section-evaluation}). The experiment results highlight the efficacy of our protocol. When ensuring a $0.9999$-probabilistic liveness, Arete is allowed to create $20$ processing shards with $500$ nodes in total whereas \textsc{GearBox} can only create 7 shards, and RIVET can create 16 shards. Remarkably, Arete can achieve $180K$ raw transactions per second ($4\times$ improvement than \textsc{GearBox} and $1.4\times$ improvement than RIVET), near intra-shard latency and $10\times$ less cross-shard latency compared to \textsc{GearBox}, and better intra-shard and cross-shard latency than RIVET.
\end{asparaitem}


\section{Preliminaries}\label{section-preliminaries}
Throughout this paper, we assume there are $n$ nodes (of which $s$ proportion are Byzantine) in the system. We consider a partially synchronous network model~\cite{DLT}, unless explicitly stated otherwise.

\subsection{BFT SMR}\label{section-preliminaries-smr}
A BFT SMR\footnote{In this work, we focus solely on the Byzantine fault model. Unless otherwise specified, we use SMR to exclusively denote BFT SMR.} is a fundamental approach in distributed computing for building Byzantine fault-tolerant systems. It can be used to commit transactions for clients with the following security properties guarantee~\cite{momose2021multi, malkhi2019flexible}:
\begin{definition}[Safety]\label{def-smr-safety}
    If sequences of transactions $(tx_1, \cdots, tx_{j})$ and $(tx'_1, \cdots, tx'_{j'})$ are committed by two honest nodes, then $tx_i=tx'_i$ for all $i\leq min\{j, j'\}$, namely, any two honest nodes commit the same prefix ledger.
\end{definition}
\begin{definition}[Liveness]\label{def-smr-liveness}
    If a transaction $tx$ is sent to at least one honest node, $tx$ will be eventually committed by all honest nodes.
\end{definition}
The fault tolerance of an SMR is constrained by a security threshold $f$, indicating the SMR ensures safety and liveness if the ratio of Byzantine nodes in the system is up to $f$.

\noindent \textbf{Implementation of SMR.} An SMR can be implemented with three repeatedly performed tasks~\cite{danezis2022narwhal}:
\begin{compactitem}[-]
    \item \textit{Data dissemination:} Nodes disperse transactions received from clients to others. This task ensures that all nodes can eventually retrieve the transaction data consistently if the disperser is honest~\cite[Definition 3.2]{nazirkhanova2022information}.
    \item \textit{Ordering:} Nodes order the dispersed transactions and output the ordered transactions into their log (namely, a tamper-proof ledger in the blockchain context). The ordering task must ensure that every honest node agrees on the same order of transactions.
    \item \textit{Execution:} After transactions are ordered, nodes execute them with an execution engine (e.g., EVM~\cite{evm} and MoveVM~\cite{movevm}). The execution results must be consistent and externally verified.
\end{compactitem}

\noindent \textbf{Fault tolerance of each task.} A BFT SMR must be designed to tolerate equivocations where Byzantine nodes vote for two conflicting blocks to break the safety property. The requirement of tolerating equivocations degrades the security threshold of SMR. For instance, it is well-known that the non-synchronous SMR can only tolerate at most a third of Byzantine nodes, i.e., $f<1/3$~\cite{DLT}. When $f<1/3$ is the fault tolerance upper bound of an SMR coupling all three SMR tasks, a decoupled SMR can achieve enhanced fault tolerance~\cite{yin2003separating}.

To elaborate, assuming the network is not synchronous, while the ordering task can tolerate only a third of Byzantine nodes, the data dissemination and execution tasks can tolerate up to half of Byzantine nodes~\cite{yin2003separating, rivet, clement2012limited, giridharan2024motorway}. 
The reason is that equivocations only exist in the ordering task. Intuitively, non-equivocation means that every node can only send the same message to different nodes in each round, in which at most one block can be committed in each round. Therefore, as long as there are more than 1/2 (honest) nodes following the protocol, the system can continuously commit blocks of transactions in the same order.
We extend them to adapt to the safety-liveness separation condition below.




\subsection{Safety-Liveness Separation in SMR}\label{section-preliminaries-sl-separation}
The classic SMR uses one security threshold $f$ to indicate both safety and liveness fault tolerances simultaneously. Some recent SMR protocols~\cite{momose2021multi, gearbox} consider the fault tolerance thresholds for safety and liveness separately to achieve a higher threshold for one property while relaxing that of another. Specifically, two thresholds can be defined when separating safety and liveness in an SMR:

\begin{definition}[Safety Threshold $f_S$]\label{def-safety-threshold}
    The safety threshold $f_S$ defines an upper bound of the ratio of Byzantine nodes that an SMR can tolerate while still ensuring safety.
\end{definition}
\begin{definition}[Liveness Threshold $f_L$]\label{def-livessness-threshold}
    The liveness threshold $f_L$ defines an upper bound of the ratio of Byzantine nodes that an SMR can tolerate while still ensuring liveness.
\end{definition}

When considering the safety-liveness separation, the fault tolerance of each SMR task can be further clarified. Specifically, $f_S$ and $f_L$ satisfy $f_S+f_L<1$ in the data dissemination and execution tasks, while $f_S+2f_L<1$ is required in the ordering task. One can prove these fault tolerance relations by setting $f_S=f_L=f$. We defer the rigorous proof to Appendix~\ref{appendix-proof-tolerance} (Lemmas~\ref{lemma-dissemination-threshold}-~\ref{lemma-execution-threshold}). Note that the definitions of safety and liveness may differ slightly for different tasks in distinct literature. For instance, safety of the data dissemination is formally defined as commitment-binding in~\cite{nazirkhanova2022information}, which similarly specifies that honest nodes commit the same block in the same round. For simplicity, we broadly use safety (\Cref{def-smr-safety}) and liveness (\Cref{def-smr-liveness}) for all SMR tasks.

\subsection{Security Properties in Sharding Systems}\label{subsection-defnitions}
The safety and liveness properties mentioned above are defined for a single blockchain/SMR.
Sharding divides nodes into groups to maintain multiple blockchains.
Followed by~\cite{zhang2023front}, we adopt the following security properties defined for a sharding system:

\begin{definition}[Sharding Safety]\label{def-system-safety}
    The sharding safety requires: (i) any two honest nodes of the same shard maintain the same prefix ledger, and (ii) any two honest nodes from two different shards have the same finalization sequence and operation (commit or abort) for all cross-shard transactions involving the two shards.
\end{definition}

\begin{definition}[Sharding Liveness]\label{def-system-liveness}
    Transactions sent to honest nodes are eventually finalized by their relevant shards.
\end{definition}


Additionally, as discussed in \S~\ref{section-introduction}, a sharding system trading liveness threshold $f_L$ for safety threshold $f_S$ might create and need to recover liveness-violated shards.
To evaluate the cost of shard recovery, we define $\mathcal{P}$-probabilistic liveness :

\begin{definition}[$\mathcal{P}$-probabilistic liveness]\label{def-system-p-probability}
    A blockchain sharding system possesses $\mathcal{P}$-probabilistic liveness if the probability of creating a new shard that guarantees liveness is at least $\mathcal{P}$.
\end{definition}

Intuitively, a higher $\mathcal{P}$-probabilistic liveness indicates a lower shard recovery cost as a newly created shard is more likely to guarantee liveness. For instance, a $0.9999$-probabilistic liveness indicates when creating 10,000 shards, there is only 1 shard violating liveness in expectation, which can meet the requirements of a commercial corporation in the real world~\cite{fournine}. It is worth emphasizing that a $\mathcal{P}$-probabilistic liveness does not signify a system guarantee of sharding liveness under a $\mathcal{P}$ probability, as the system can recover temporary liveness-violated shards via the reconfiguration mechanism (see \S~\ref{subsection-system-architecture} for more details). Therefore, a sharding system with a $\mathcal{P}$-probabilistic liveness can still guarantee the same level of security as previous sharding protocols~\cite{Rapidchain, Omniledger, gearbox}.

\section{Building A Scalable Sharding System with Optimal-size Shards}
\label{section-problem} 
A blockchain sharding system divides nodes into multiple shards to handle transactions in parallel. Intuitively, with more lightweight shards created, the sharding system has higher parallelism and becomes more scalable. This section will present how to enable the system to create more shards by reducing the shard size.

\subsection{Calculate the Minimum Shard Size}
We assume a sharding system has a total population of $n$ nodes where at most $s$ fraction of the nodes are Byzantine. Like prior works~\cite{Omniledger, Rapidchain, gearbox}, we can employ the hypergeometric distribution to calculate the probability of forming an insecure shard with size $m$. Specifically, denote $\mathit{FAU}$ as the event where the insecure shard contains more than $f$ Byzantine nodes, with $f$ representing the security threshold. The probability of $\mathit{FAU}$ happening is:

\begin{equation}
Pr[\mathit{FAU}]=\sum_{x=\lceil mf \rceil}^{m}\frac{\binom{ns}{x}\binom{n-ns}{m-x}}{\binom{n}{m}}.
\label{eq-formation}
\end{equation}

\noindent We hope to keep the probability of forming faulty shards negligible, such that a shard can guarantee its security properties under a high-security level. In particular, we define a security parameter $\lambda$ and hope to satisfy the following formulation:
\begin{equation}
Pr[\mathit{FAU}]_{m=m^*} \leq 2^{-\lambda}.
\label{eq-mini}	
\end{equation} 
\noindent Equation (\ref{eq-mini}) can be used to evaluate the minimum shard size $m^*$ that enables shards to ensure security properties with a high probability, i.e., $\lambda$-bit security. Specifically, given $n$, $s$, $f$, and $\lambda$, the minimum shard size $m^*$ is the minimum $m$ that satisfies Equation (\ref{eq-mini}). 
\subsection{Existing Solutions and Our Key Insights}\label{section-comparison-insight}
Equations (\ref{eq-formation})-(\ref{eq-mini}) indicate the relationships among $n$, $\lambda$, $s$, $f$, and $m^*$. We focus on the impacts of $s$ and $f$ on $m^*$ as distinct sharding protocols can have the same $n$ and $\lambda$ but assume varying $s$ and $f$ to reduce $m^*$ (we also provide an overall analysis for interested readers in Appendix~\ref{appendix-size-evaluation}).
Shortly, we can conclude that a smaller $s$ or a larger $f$ allows the system to derive a smaller $m^*$. This insight has actually been noted by many representative sharding protocols. We compare their solutions for reducing shard size in Table~\ref{table:related work}. Here, $k$ is the number of shards that can be created, and $Pr[f_L]$ denotes the probability that no more than an $f_L$ fraction of Byzantine nodes are assigned to a shard. To elaborate further, $Pr[f_L]$ indicates the probability that the liveness of a newly created shard is guaranteed, which equals $Pr[f_L]$-probabilistic liveness as defined in Definition~\ref{def-system-p-probability}.


To reduce shard size, many protocols~\cite{Omniledger, Rapidchain, ahl} make stronger assumptions. Omniledger~\cite{Omniledger} assumes a smaller $s$ to reduce $m^*$. RapidChain~\cite{Rapidchain} assumes a synchronous network by which $f$ can be increased to $49\%$ via a synchronous BFT protocol. AHL~\cite{ahl} relies on trusted hardware in the intra-shard consensus such that the consensus leader cannot equivocate, and thus can increase $f$ to $49\%$. \textsc{GearBox}~\cite{gearbox} separates the safety and liveness of a shard and trades liveness threshold $f_L$ for safety threshold $f_S$ to get a larger $f_S$. To elaborate, \textsc{GearBox} replaces $f$ of Equation (\ref{eq-formation}) with a larger $f_S$ to get a smaller $m^*$, by which a newly created shard statistically ensures the safety but may violate the liveness due to $f_S>f_L$. Thus, \textsc{GearBox} only provides a $\mathcal{P}$-probabilistic liveness. All the above sharding protocols employ every shard to perform all tasks of an SMR. The overall fault tolerance of their shards $f$ is limited by the lower-bound fault tolerance required by the ordering task, e.g., $f<1/3$ in~\cite{Omniledger, gearbox} that do not assume a synchrony network or rely on trusted hardware. Recall from \S~\ref{section-preliminaries-sl-separation} that a shard performing the ordering task with $f<1/3$ threshold necessitates $f_S+2f_L<1$. Due to this limitation, \textsc{GearBox} only increases $f_S$ slightly if the system aims to provide a high $\mathcal{P}$-probabilistic liveness. For instance, to achieve $0.9999$-probabilistic liveness with $s=25\%$, $f_S$ can be set to at most $35\%$, allowing only two shards to be created even with the safety-liveness separation mechanism.

To free shards from the limitation of the ordering fault tolerance, RIVET~\cite{rivet} proposes a reference-worker sharding scheme. In RIVET, the whole sharding system implements one SMR, and only one reference shard runs the ordering task. The worker shards exclusively disseminate and execute transactions and thus can tolerate $49\%$ Byzantine nodes. Our Arete follows such a sharding scheme but employs new mechanisms to further reduce shard size and enable efficient transaction handling (see more comparison in \S~\ref{main-body-related-work}).

\noindent \textbf{Key insights in creating optimal-size shards.} This paper aims to reduce the shard size to create more lightweight shards without making stronger assumptions. To create optimal-size shards, a key insight is to enable a larger fault tolerance threshold of shards. This can be done by freeing a shard from the limitation of the ordering fault tolerance and separating safety and liveness (but needed to avoid a high cost of shard recovery as in \textsc{GearBox}).




\begin{table}[t]
	\centering
	\footnotesize
	\caption{Comparison of sharding protocols on the total Byzantine ratio $s$, safety threshold $f_S$, liveness threshold $f_L$, minimum shard size $m^*$, shard number $k$, ensured $Pr[f_L]$-probabilistic liveness, and network assumption. Assume $n=1000$ nodes in the system and a $\lambda=30$ security parameter.}
	\label{table:related work}
	\begin{threeparttable}
	\begin{tabular}
 {>{\centering\arraybackslash}p{0.095\textwidth}|>{\centering\arraybackslash}p{0.02\textwidth}|>{\centering\arraybackslash}p{0.02\textwidth}|>{\centering\arraybackslash}p{0.02\textwidth}|>{\centering\arraybackslash}p{0.015\textwidth}|>{\centering\arraybackslash}p{0.01\textwidth}|>{\centering\arraybackslash}p{0.055\textwidth}|c}
	\hline 
	Protocol& $s$ & $f_S$ & $f_L$ & $m^*$ & $k$ & $Pr[f_L]$ & \makecell{network \\ assumption}  \\
	\hline
	Omniledger~\cite{Omniledger}& $25\%$ & $33\%$ & $33\%$ & 486 & 2 & $1-2^{-30}$ & partial sync. \\
	RapidChain~\cite{Rapidchain}& $33\%$ & $49\%$ & $49\%$ & 247 & 4 & $1-2^{-30}$ & sync. \\
	AHL~\cite{ahl}& $30\%$& $49\%$& $49\%$ & 182 & 5 & $1-2^{-30}$ & partial sync. \\
	RIVET~\cite{rivet}& $33\%$& $49\%$& $49\%$ & 247 & 4 & $1-2^{-30}$ & partial sync.\\
	
	\hline
	\multirow{3}*{\textsc{GearBox}~\cite{gearbox} }
	& $30\%$ & $39\%$& $30\%$ & 477 & 2 & $0.5768$ & \multirow{3}*{partial sync.} \\
        & $25\%$ & $35\%$ & $32\%$ & 403 & 2 & $0.9999$ \\
	& $25\%$ & $57\%$ & $21\%$ & \textcolor{ForestGreen}{72} &  \textcolor{ForestGreen}{13} & \textcolor{Red}{$0.3422$} \\
	\hline
	\multirow{2} * {Arete (ours)} 
	& $30\%$ & $54\%$ & $45\%$ & \textcolor{ForestGreen}{125} & \textcolor{ForestGreen}{8} & \textcolor{ForestGreen}{$0.9999$} & \multirow{2}*{partial sync.} \\
	& $25\%$ & $57\%$ & $42\%$ & \textcolor{ForestGreen}{72} & \textcolor{ForestGreen}{13} & \textcolor{ForestGreen}{$0.9999$} \\
	\hline
	\end{tabular}


	\end{threeparttable}
\vspace{-5mm}
\end{table}

\section{Arete Protocol}\label{section-overview}
This section introduces our highly scalable sharding protocol Arete.
\subsection{System Setting and Trust Assumption}\label{subsection-model}
\noindent \textbf{Transaction model.} We target the account transaction model because it is more powerful to support different functionalities with smart contracts.
An intra-shard transaction is defined as a transaction that involves states/data from one shard. In contrast, a cross-shard transaction involves states from several shards. In this paper, we focus on the \textit{one-shot} cross-shard transactions, such as cross-shard transfer and atomic swap transactions, which account for the majority in a realistic network~\cite{COSPLIT}. 
A one-shot transaction can be divided into two intra-shard transactions for relevant shards to execute independently without exchanging involved states (see Appendix~\ref{appendix-comparison} for more details). Supporting \textit{multi-shot} cross-shard transactions is still an open problem for all existing sharding protocols because of the state separation. Some potential solutions built on our sharding protocol will be discussed in Appendix~\ref{appendix-extension-details}.

\noindent \textbf{Threat model.} 
There are $n$ nodes
(of which $s$ proportion are Byzantine) in the system. A Byzantine (malicious) node can behave arbitrarily to deviate from the sharding protocol. However, Byzantine nodes are computationally bounded and cannot break standard cryptographic constructions. Based on these assumptions, we say a shard guarantees safety (or liveness) statistically if the probability of forming the shard with no more than $f_S$ (of $f_L$) ratio of Byzantine nodes satisfies a $\lambda$-bit security level (i.e., the probability is more than $1-2^{-\lambda}$).
Additionally, like all sharding protocols, our protocol operates in \textit{epochs}, which can be defined by a fixed number of blocks (e.g., 1000 blocks). At the beginning of each epoch, nodes reorganize shards through a reconfiguration mechanism to prevent adaptive attacks~\cite{hybrid}. As \textsc{GearBox}~\cite{gearbox}, the reconfiguration mechanism is also used to recover liveness-violated processing shards (\S~\ref{subsection-system-architecture}).

\noindent \textbf{Network model.} As our ordering shard runs a partially synchronous BFT protocol (i.e., Hotstuff~\cite{hotstuff}), we assume a partially synchronous network model, where there is an unknown global stabilization time (GST) after which all message delays between honest nodes are bounded by a known $\Delta$~\cite{DLT}.



\subsection{Generating Optimal-size Shards with Arete}\label{subsection-system-architecture}
\begin{figure}[t]
    \setlength\abovecaptionskip{-0.1\baselineskip}
    \setlength\belowcaptionskip{-1.0\baselineskip}  
    \centering
    \includegraphics[width=3.3in]{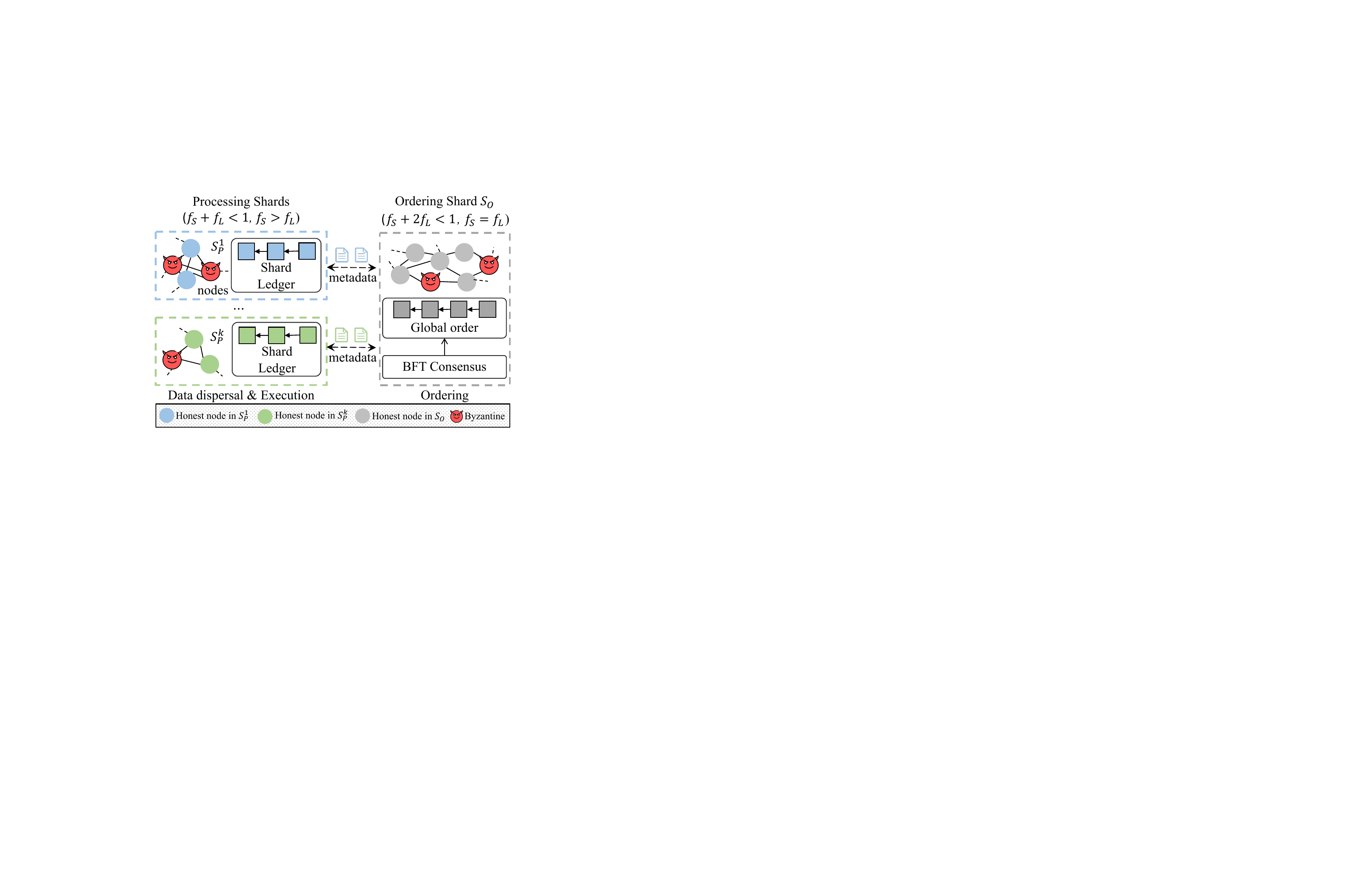}
    \caption{Arete overview: the system is divided into one ordering shard and $k$ processing shards $\{S_P^1 \cdots S_P^{k}\}$. The ordering shard runs a BFT consensus to globally order transactions, tolerating up to $f_S=f_L<1/3$ Byzantine nodes. A processing shard performs the data dissemination and execution tasks, tolerating up to $f_S\geq1/2$ Byzantine nodes. }
    \label{fig:architecture3}
    \vspace{-3mm}
\end{figure}

Arete is designed to achieve optimal shard size and is considered an optimal sharding protocol. By ``optimal", we mean that this is the minimum shard size a sharding can achieve under the same system parameters, trust and network assumptions. The key idea is to allow a larger safety threshold $f_S$. To achieve this, Arete consists of two building blocks: ordering-processing sharding scheme and safety-liveness separation. Figure~\ref{fig:architecture3} gives an overview of Arete.

\noindent\textbf{(1) Ordering-processing sharding scheme}. In Arete, there are two types of shards: \textit{one} ordering shard and \textit{multiple} processing shards. An ordering shard performs the ordering task of SMR. It runs a BFT consensus protocol to establish a \textit{global} transaction order for the system. Processing shards perform the data dissemination and execution tasks of SMR. They work as the ledger maintainer and transaction executor. 

On a high level, our sharding protocol \textit{shards ledger maintenance and transaction execution but not consensus}. 
By decoupling transaction processing (i.e., dispersing and executing) from ordering, Arete enables processing shards to be free from consensus. When there is no equivocation, each processing shard can tolerate up to $f<1/2$ ratio of Byzantine nodes even under a non-synchronous network model~\cite{yin2003separating, clement2012limited}. The improved fault tolerance threshold allows a smaller shard size and the creation of more shards as we analyzed in \S~\ref{section-problem}. Note that the ordering shard running the consensus protocol is designed to statistically guarantee safety and liveness and thus still requires a security threshold $f=f_S=f_L<1/3$.

\noindent\textbf{(2) Safety-liveness separation}. Arete considers safety and liveness against Byzantine nodes separately in processing shards. Like \textsc{GearBox}, Arete guarantees the safety property of processing shards statistically, while allowing a processing shard to violate liveness for a while. Specifically, when forming processing shards, Arete uses the safety threshold $f_S$ to compute the minimum shard size, ensuring that the proportion of malicious nodes in each processing shard does not exceed $f_S$. By appropriately increasing the safety threshold $f_S$ and decreasing the liveness threshold $f_L$, Arete enables a newly created processing shard with optimal size to guarantee safety statistically. Unlike \textsc{GearBox}, Arete is designed to ensure a high $\mathcal{P}$-probabilistic liveness (thus a low shard recovery cost). Specifically, recall from \S~\ref{section-preliminaries-sl-separation} that the safety-liveness separation enables our processing shards that only perform data dissemination and execution tasks to have $f_S+f_L<1$. This is a significant improvement compared to the previous safety-liveness separation scheme~\cite{gearbox, instachain} which necessitate $f_S+2f_L<1$.
As a result, when setting $f_S=57\%$ to reduce the shard size to $m=72$, Arete allows $f_L=42\%$ (due to $f_S+f_L<1$) and a relatively high $0.9999$-probabilistic liveness (due to a high $f_L=42\%$), as shown in Table~\ref{table:related work}.




\noindent \textbf{Shards bootstrap}. 
The process of bootstrapping Arete closely follows that of previous sharding protocols using the random assignment mechanism~\cite{gearbox, Omniledger}. 
We consider a permissionless setting where nodes are allowed to participate in the protocol with an identity obtained from permissionless Sybil-attack-resistant foundations (e.g., Proof-of-Stake). We assume that the identities of all nodes are public (e.g., via publishing identities on an existing blockchain or publicly available websites). Given the total number of nodes $n$, the security parameter $\lambda$, the assumed total ratio of Byzantine nodes $s$, and the required liveness threshold $f_L$, nodes can form shards with the shard size $m$ calculated through Equation~(\ref{eq-formation}) and (\ref{eq-mini}). Note that the shard sizes of the ordering shard (denoted by $m^\#$) and processing shards (denoted by $m^*$) are different, where $m^\# > m^*$. The ordering shard must be large enough to guarantee both safety and liveness properties statistically, where $f=f_S=f_L=1/3$. In contrast, a processing shard can have fewer nodes, ensuring safety statistically but only providing $0.9999$-probabilitic liveness. The formations of the ordering shard and processing shards are independent.
Specifically, every node calculates a hash by concatenating its identity and randomness~\cite{Elastico, Omniledger, gearbox}. Nodes first form the ordering shard based on the ranking of their calculated hash values (e.g., with the $m^\#$ largest nodes forming an $m^\#$-size ordering shard). After that, the resulting hash values of the left nodes are mapped to a range $[0, 1)$, which is partitioned into $k=n/m^*$ regions\footnote{For simplicity, we ignore the reduction of $n$ caused by the formation of the ordering shard because $m^\#$ is much smaller than $n$ in practice where tens of thousands of nodes exist in the network. For instance, $m^\#/n \approx 1\%$ with $n=100,000$, $\lambda=30$, and $s=25\%$. We can also achieve this by enabling shard overlap as previous sharding protocols do~\cite{Elastico, Rapidchain, gearbox}.} to identify processing shards to which nodes belong. Note that the bootstrapping procedure in our work follows the random assignment mechanism as in previous works, and thus it can form (both ordering and processing) shards randomly.


\noindent \textbf{Shards recover.} Despite the potential for Arete to generate liveness-violated processing shards due to its safety-liveness separation, the shard reconfiguration mechanism comes into play to recover these liveness-violated processing shards. Our reconfiguration mechanism follows the methodology employed by \textsc{GearBox} and therefore inherently implies its effectiveness and security. 
Specifically, if the ordering shard receives messages (see \S~\ref{sec-coe-model} for more details about the messages) from less than $f_S\cdot|S_P^{sid}|+1$ nodes of a processing shard $S_P^{sid}$ during an epoch, it can assign more nodes to $S_P^{sid}$, setting a higher liveness threshold (thus resulting in a lower safety threshold and a larger shard size). 
This process is iterated until the ordering shard receives messages from enough nodes of the processing shard, but when entering a new epoch, the liveness-recovered processing shard will be trimmed to the originally optimal shard size. 
During the recovery process, the newly assigned nodes need to synchronize the history data of the processing shard. However, Arete makes sure this recovery procedure rarely happens, e.g., $0.01\%$ probability as we guarantee $0.9999$-probabilistic liveness. To see how this high-probabilistic liveness reduces recovery costs, consider Ethereum state synchronization. Currently, a node can synchronize 450 GB of Ethereum states in around 12 hours~\cite{eth-sync}. The expected recovery cost for Arete, with $0.9999$-probabilistic liveness, is $12\times0.0001=0.0012$ hours. In contrast, \textsc{GearBox}, which guarantees $0.3422$-probabilistic liveness (\Cref{table:related work}), requires $12\times0.6578=7.8936$ hours for recovery, resulting in 6578 times higher recovery costs than Arete.

\section{The COE Architecture}\label{sec-coe-model}
In this section, we present a new certify-order-execute (COE) architecture that makes Arete handle transactions efficiently. 

\subsection{Overview}\label{sec-coe-model-overview}
Similar to recent high-performance blockchain protocols~\cite{danezis2022narwhal, spiegelman2022bullshark, giridharan2024motorway, duan2024dashing}, the COE architecture in Arete handles transactions in three decoupled stages: disseminating transactions in the \text{\footnotesize CERTIFY} stage, ordering transactions in the \text{\footnotesize ORDER} stage, and executing transactions in the \text{\footnotesize EXECUTE} stage. However, unlike these protocols, Arete allows multiple processing shards to disseminate (and execute) transactions in parallel while a single ordering shard orders transactions.

\begin{figure}
    \setlength\abovecaptionskip{0.1\baselineskip}
    \setlength\belowcaptionskip{-1.0\baselineskip} 
    \centering
    \includegraphics[width=3.3in]{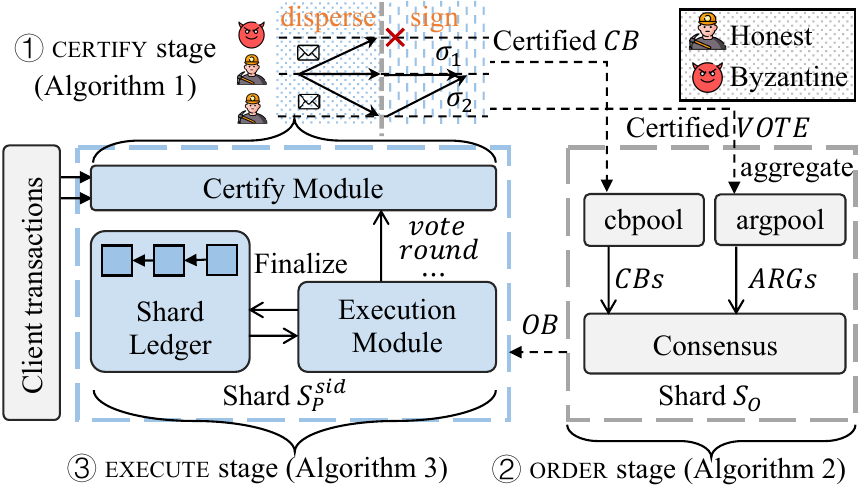}
    \caption{The COE architecture overview: \ding{172} The \text{\footnotesize CERTIFY} stage: each processing shard disseminates and generates certified messages, performing the data dissemination task. \ding{173} The \text{\footnotesize ORDER} stage: the ordering shard performs the ordering task to establish a global order. \ding{174} The \text{\footnotesize EXECUTE} stage: each processing shard executes and finalizes the ordered transactions.}
    \label{fig:coe-nutshell}
    \vspace{-2mm}
\end{figure}

Figure~\ref{fig:coe-nutshell} shows the overview of the COE architecture. To elaborate, each processing shard disseminates transactions within the shard itself, forming \textit{certificates} of these disseminated transactions (the \text{\footnotesize CERTIFY} stage \ding{172}). The certificates consist of a quorum of signatures from nodes in the processing shard, ensuring that at least one honest node possesses intact transactions. The ordering shard runs a consensus protocol to order these certified transactions, ensuring every honest node has a consistent order for the relevant transactions (the \text{\footnotesize ORDER} stage \ding{173}). The ordered transactions are then executed by processing shards consistently (the \text{\footnotesize EXECUTE} stage \ding{174}). 

To understand how the COE architecture works efficiently, notice that intact transactions are disseminated only during the \text{\footnotesize CERTIFY} stage, while the \text{\footnotesize ORDER} stage only needs to order extremely lightweight metadata (i.e., signed-hash digests), and \text{\footnotesize EXECUTE} stage is performed locally without involving transaction dissemination. This is because the certificates generated in the \text{\footnotesize CERTIFY} stage guarantee data (transactions) availability such that 1) nodes in the ordering shard can verify the validity of ordered transactions without possessing them, and 2) nodes in processing shard can retrieve the intact transactions for execution with the certified metadata. We elaborate on each stage below.

\subsection{Architecture Specification}\label{sec-coe-specification} 
In Arete, nodes perform data dissemination, ordering, and execution tasks in the \text{\footnotesize CERTIFY}, \text{\footnotesize ORDER}, and \text{\footnotesize EXECUTE} stages respectively.

\textbf{(1) \text{\footnotesize CERTIFY} stage} (Algorithm~\ref{alg-execution-node-certify}). 
In this stage, nodes in every processing shard disseminate intact transactions within the shard and generate many lightweight certificates that are then forwarded to the ordering shard for ordering. 
Specifically, the \text{\footnotesize CERTIFY} stage is performed in a leaderless manner. Each node $\mathrm{N}_j^{\mathit{sid}}$ in a processing shard $S_P^{\mathit{sid}}$ can create a new \textit{execution block} $\mathit{EB} = \langle \mathit{itxs}, \mathit{ctxs}, \mathit{sid}, j \rangle$ with intra-shard transactions $\mathit{itxs}$ and cross-shard transactions $\mathit{ctxs}$ directly received from clients (such a direct assignment can be achieved easily based on the addresses of a transaction's sender and receiver), and then broadcast $\mathit{EB}$ attached with its signature $\sigma_j$ (lines 1-4).
Once a node receives $\mathit{EB}$, it helps to certify that it has received intact transaction data of $\mathit{EB}$ (lines 6-9). Specifically, it signs a \text{\scriptsize CERTIFIED} message $\mathit{ct}$ that contains the digest of $\mathit{EB_{dst}}$, the metadata of cross-shard transactions in $\mathit{M_{ctxs}}$, and the shard ID $sid$, and sends the signed $\mathit{ct}$ back to the sender node. The creator of $\mathit{EB}$ (i.e., $\mathrm{N}_j^{\mathit{sid}}$ in this example) can then use these signed messages to create a certificate block $\mathit{CB} = \langle \mathit{EB_{dst}}, \mathit{M_{ctxs}}, \mathit{sid}, \mathit{\sigma^{P}_{set}} \rangle$ that contains $f_S\cdot |S_P^{\mathit{sid}}|+1$ signatures $\mathit{\sigma^{P}_{set}}$, and send $\mathit{CB}$ to the ordering shard (lines 10-14). Since the quorum $f_S\cdot |S_P^{\mathit{sid}}|+1$ of signatures ensures the data availability of transactions (see Lemma~\ref{lemma-data-availability} below), the ordering shard now can use the more lightweight certificate blocks rather than execution blocks for ordering.


\begin{algorithm}[t]
    \footnotesize
    \caption{\text{\footnotesize CERTIFY} stage for Node $\mathrm{N}_j^{\mathit{sid}}$ in shard $S^{\mathit{sid}}_P$}
    \label{alg-execution-node-certify}
    \begin{algorithmic}[1]
        \Statex {\footnotesize \color{Peach}$\blacktriangleright$  certify new transactions}
        \State \textbf{upon} certify(transactions $\mathit{txs}$) \textbf{do}
        \\\Comment{{\footnotesize \color{gray} separate $\mathit{txs}$ into intra-shard transactions $\mathit{itxs}$ and cross-shard transactions $\mathit{ctxs}$ based on their involved processing shards}}
        \State \hspace*{4mm} $\mathit{EB} \leftarrow \langle \mathit{itxs}, \mathit{ctxs}, \mathit{sid}, j \rangle$
        \State \hspace*{4mm} disseminate $\langle$\text{\scriptsize CERTIFY}, $\mathit{EB} \rangle_{\sigma_j}$ to $S^{\mathit{sid}}_P$
        \State \hspace*{4mm} $\mathit{certified_{EB}} \leftarrow \{\}$
        
        \State \textbf{upon} receiving $\langle$\text{\scriptsize CERTIFY}, $\mathit{EB}\rangle_{\sigma_k}$ \textbf{do}
        \State \hspace*{4mm} $\mathit{EB_{dst}} \leftarrow$ compute $\mathit{EB}$'s digest
        \State \hspace*{4mm} $\mathit{M_{ctxs}} \leftarrow$ $\{$getMetadata($\mathit{ctx}$) $| \forall \mathit{ctx} \in \mathit{EB.ctxs}\}$
        \State \hspace*{4mm} send $ct=\langle$\text{\scriptsize CERTIFIED}, $\mathit{EB_{dst}}, \mathit{M_{ctxs}}, \mathit{sid}\rangle_{\sigma_j}$ to Node $k$

        \State \textbf{upon} receiving a \text{\scriptsize CERTIFIED} message $\mathit{ct}$ \textbf{do}
        \State \hspace*{4mm} $\mathit{certified_{EB}} \leftarrow \mathit{certified_{EB}} \cup \{\mathit{ct}\} $
        \State \hspace*{4mm} \textbf{if} $|\mathit{certified_{EB}}| \geq f_S\cdot |S^{\mathit{sid}}_P|$ \textbf{then}
        \State \hspace*{8mm} $\mathit{CB} \leftarrow \langle \mathit{EB_{dst}}, \mathit{M_{ctxs}}, \mathit{sid}, \sigma_j\cup\{\mathit{certified_{EB}.\sigma}\} \rangle$
        \State \hspace*{8mm} send $\mathit{CB}$ to the ordering shard
        
        \Statex {\footnotesize \color{Peach}$\blacktriangleright$  certify the ordered cross-shard transactions}
        \State \textbf{upon} receiving a \text{\scriptsize EXEVOTE} message $\mathit{ev}$ from its execution module \textbf{do}
        \State \hspace*{4mm} disseminate $\langle$\text{\scriptsize CVOTE}, $\mathit{ev.vote}, \mathit{ev.round}, \mathit{sid}, j \rangle_{\sigma_j}$ to $S^{\mathit{sid}}_P$
        \State \hspace*{4mm} $\mathit{voted} \leftarrow \{\}$

        \State \textbf{upon} receiving $\langle$\text{\scriptsize CVOTE}, $\mathit{vote}, \mathit{round}, \mathit{sid}, k \rangle_{\sigma_k}$ \textbf{do}
        \State \hspace*{4mm} \textbf{if} $\mathit{vote}$ is correct \textbf{then}
        \State \hspace*{8mm} send $\langle$\text{\scriptsize CVOTED}, $\mathit{vote}, \mathit{round}, \mathit{sid} \rangle_{\sigma_j}$ to Node $k$

        \State \textbf{upon} receiving a \text{\scriptsize CVOTED} message $\mathit{vt}$ \textbf{do}
        \State \hspace*{4mm} $\mathit{voted} \leftarrow \mathit{voted} \cup \{vt\} $
        \State \hspace*{4mm} \textbf{if} $|\mathit{voted}| \geq f_S\cdot |S^{\mathit{sid}}_P|$ \textbf{then}
        \State \hspace*{8mm} $\mathit{VOTE} \leftarrow \langle \mathit{vote}, \mathit{round}, \mathit{sid}, \sigma_j\cup\{\mathit{voted.\sigma}\} \rangle$
        \State \hspace*{8mm} send $\mathit{VOTE}$ to the ordering shard
    \end{algorithmic}
\end{algorithm}

\begin{lemma}[Data Availability]\label{lemma-data-availability}
    Given a processing shard with $|S_P^{\mathit{sid}}|$ nodes and a safety threshold $f_S$, any honest node can recover an intact execution block $\mathit{EB_1}$ if $\mathit{EB_1}$'s associated certificate block contains at least $f_S\cdot |S_P^{\mathit{sid}}|+1$ signatures from distinct nodes of $S_P^{\mathit{sid}}$.
\end{lemma}

\begin{myproof}{}
    Since Arete ensures that there are no more than $f_S\cdot |S_P^{\mathit{sid}}|$ malicious nodes in each processing shard, the quorum (i.e., $f_S\cdot |S_P^{\mathit{sid}}|+1$) of signatures ensures that at least one honest node has the intact execution block. This enables all nodes in $S_P^{\mathit{sid}}$ to eventually obtain the block by synchronizing it from the honest node, ensuring data availability.
\end{myproof}

Since blockchain sharding introduces cross-shard transactions, nodes in a processing shard also need to certify the $\mathit{VOTE}$ results of cross-shard transactions (lines 15-25). A $\mathit{VOTE}$ result is a data structure mapping cross-shard transactions to the locally-executed results (i.e., success or failure) and will be aggregated by the ordering shard to determine if its corresponding cross-shard transactions can be committed (see the \text{\footnotesize ORDER} stage below). Specifically, upon receiving a \text{\scriptsize EXEVOTE} message from its execution module, $\mathrm{N}_j^{\mathit{sid}}$ disseminates its $\mathit{VOTE}$ result and collects signatures from other nodes in $S_P^{\mathit{sid}}$. Similarly, the node then sends the $\mathit{VOTE}$ result to the ordering shard after it collects at least $f_S\cdot |S_P^{\mathit{sid}}|+1$ signatures. The quorum of signatures ensures that at least one honest node certifies the correctness of the $\mathit{VOTE}$ result. Therefore, anyone (in the ordering shard) can verify the correctness of a $\mathit{VOTE}$ result without re-executing the corresponding transactions. 


\begin{algorithm}[t]
    \footnotesize
    \caption{\text{\footnotesize ORDER} stage for the ordering shard $S_{O}$}
    \label{alg-ordering-node-order}
    \begin{algorithmic}[1]
        \Statex \textbf{Local data}: $\mathit{voteRounds}$  \Comment{{\footnotesize \color{gray} map processing shard IDs to the latest consensus rounds of $\mathit{VOTE}$ results received from processing shards} } 

        \State \textbf{upon} receiving a $\mathit{CB}$ from shard $S_P^{\mathit{sid}}$  \textbf{do}
        \State \hspace*{4mm} \textbf{if} Verify($\mathit{CB}, \mathit{CB.\sigma^P_{set}}$) \textbf{then}
        \State \hspace*{8mm} store $\mathit{CB}$ into cbpool
        
        \Statex {\footnotesize \color{Peach}$\blacktriangleright$ aggregate $\mathit{VOTE}$ results for cross-shard transactions}
        \State \textbf{upon} receiving a $\mathit{VOTE}$ result  \textbf{do}
        \State \hspace*{4mm} \textbf{if} Verify($\mathit{VOTE}, \mathit{VOTE.\sigma^P_{set}}$) \textbf{then}
        \State \hspace*{8mm} \textbf{if} $\mathit{voteRounds}[\mathit{VOTE.sid}] + 1 = \mathit{VOTE.round}$ \textbf{then}
        \State \hspace*{12mm} $\mathit{voteRounds}[\mathit{VOTE.sid}] \leftarrow$ $\mathit{VOTE.round}$
        \State \hspace*{12mm} $\mathit{ARG} \leftarrow$ fetch aggregator of $\mathit{VOTE.round}$
        \State \hspace*{12mm} $\mathit{ARG} \leftarrow$ \textsc{AggreateVote}($\mathit{ARG}$, $\mathit{VOTE.vote}$)
        \State \hspace*{12mm} \textbf{if} $\mathit{ARG}$ is ready to be finalized \textbf{then}
        \State \hspace*{16mm} store $\mathit{ARG}$ into argpool 
        \State \hspace*{8mm} \textbf{elseIf} $\mathit{voteRounds}[\mathit{VOTE.sid}] < \mathit{VOTE.round}$ \textbf{then}
        \State \hspace*{12mm} synchronize missing votes from $S^{\mathit{VOTE.sid}}_P$
        \State \hspace*{12mm} update the relevant aggregators

         \Statex {\footnotesize \color{Peach}$\blacktriangleright$  order new transactions}
        \State \textbf{when} creating a new ordering block $\mathit{OB}$ \textbf{do}
        \State \hspace*{4mm} $\mathit{ARGs} \leftarrow$ fetch aggregated votes from its argpool
        \State \hspace*{4mm} $\mathit{CBs} \leftarrow$ fetch certificate blocks from its cbpool
        \State \hspace*{4mm} $\mathit{OB} \leftarrow$ createBlock($newRound$, $\mathit{ARGs}$, $\mathit{CBs}$) 
        \State \hspace*{4mm} coordinate a consensus instance for $\mathit{OB}$
        
        \Statex {\footnotesize \color{Peach}$\blacktriangleright$ inform processing shards}
        \State \textbf{upon} finalizing the new ordering block $\mathit{OB}$ \textbf{do}
        \State \hspace*{4mm} send $\mathit{OB}= \langle \mathit{ARGs}, \mathit{\hat{EB}_{dsts}}, \mathit{\hat{M}_{ctxs}}, r, \mathit{\sigma^{O}_{set}}\rangle$ to processing shards
        
        \Statex
        \State \textbf{function} \textsc{AggreateVote}($\mathit{ARG}$, $\mathit{vote}$)
        \State \hspace*{4mm} \textbf{for} $\forall \mathit{M_{ctx}} \in \mathit{vote.key()}$ \textbf{do}
        \State \hspace*{8mm} \textbf{if} $\mathit{ARG.vote}$.contains($\mathit{M_{ctx}}$) \textbf{then}
        \State \hspace*{12mm} $\mathit{ARG.vote}[\mathit{M_{ctx}}] \leftarrow \mathit{ARG.vote}[\mathit{M_{ctx}}]$ $\&\&$ $\mathit{vote}[\mathit{M_{ctx}}]$
        \State \hspace*{8mm} \textbf{else}
        \State \hspace*{12mm} $\mathit{ARG.vote}[\mathit{M_{ctx}}] \leftarrow \mathit{vote}[\mathit{M_{ctx}}]$
        \State \hspace*{4mm} \textbf{return} $\mathit{ARG}$
    \end{algorithmic}
\end{algorithm}

\textbf{(2) \text{\footnotesize ORDER} stage} (Algorithm~\ref{alg-ordering-node-order}). The \text{\footnotesize ORDER} stage then helps globally order transactions indicated by certificate blocks from processing shards. The COE architecture orders transactions round-by-round. In each consensus round $r$, the ordering shard $S_O$ runs an instance of consensus to create and finalize an \textit{ordering block} $\mathit{OB}$. 

Since Arete is agnostic to the consensus protocol, we omit the procedure of consensus on finalizing the ordering block in Algorithm~\ref{alg-ordering-node-order}, which depends on the specific consensus protocol. For instance, when running Hotstuff~\cite{hotstuff} in $S_O$, a leader $\mathcal{L}$ is designated to create $\mathit{OB}$ (lines 15-19). Specifically, $\mathcal{L}$ fetches and orders a set of certificate blocks from its certificate pool cbpool, where cbpool stores all valid certificate blocks received from processing shards (lines 1-3). Note that a certificate block is considered valid if it contains a quorum of (i.e., at least $f_S\cdot|S_P^{\mathit{sid}}|+1$) signatures from its processing shard $S_P^{\mathit{sid}}$. Once the fetched certificate blocks $\mathit{CBs}$ are ordered, the local orders of transactions inside $\mathit{CBs}$ are retained unchanged and combined into a global order. Then $\mathcal{L}$ coordinates several stages of Hotstuff to finalize $\mathit{OB}$.

Arete also assigns the ordering shard to coordinate the finalization of cross-shard transactions.
Arete adopts an \textit{asynchronous cross-shard commit approach}, where the ordering shard can move to the next consensus round without waiting to receive $\mathit{VOTE}$ results of the current round from all relevant processing shards, i.e., cross-shard transactions are finalized in a non-blocking way. To this end, nodes trace the latest consensus round of the $\mathit{VOTE}$ result from each processing shard, i.e., $\mathit{voteRounds}$. When receiving a certified $\mathit{VOTE}$ result, nodes aggregate it into an aggregator $\mathit{ARG}$ that is associated with a consensus round $\mathit{ARG.round}$ (lines 4-14).

The aggregator $\mathit{ARG}$ will serve as a reference to indicate if a cross-shard transaction can be committed. Briefly speaking, the aggregation process \textsc{AggregateVote} (lines 22-28) is to perform AND operation on the execution results (where 1 means success and 0 means failure) of cross-shard transactions from all relevant processing shards. If the final value corresponding to a cross-shard transaction in an aggregator $\mathit{ARG}$ is 1, then the cross-shard transaction is executed successfully by all relevant processing shards; in contrast, 0 indicates at least one processing shard fails to execute the cross-shard transaction. Once a node receives all involved $\mathit{VOTE}$ results for a previous consensus round $\mathit{VOTE.round}$, the corresponding aggregator $\mathit{ARG}$ is ready to be finalized and stored in the node's aggregator pool argpool (Line 11). When creating a new ordering block $\mathit{OB}$, the leader $\mathcal{L}$ also fetches aggregators from argpool and orders them in the block based on their corresponding consensus rounds (Line 16). The new ordering block $\mathit{OB}= \langle \mathit{ARGs}, \mathit{\hat{EB}_{dsts}}, \mathit{\hat{M}_{ctxs}}, r, \mathit{\sigma^{O}_{set}}\rangle$ is then sent to processing shards for execution (Lines 20-21), where $\mathit{ARGs}$ is a set of aggregations of execution results for cross-shard transactions, $\mathit{\hat{EB}_{dsts}}$ is a list of execution block digests, $\mathit{\hat{M}_{ctxs}}$ is a list of cross-shard transactions metadata, and $\mathit{\sigma^{O}_{set}}$ is a quorum (i.e., $2f\cdot|S_O|+1$) of signatures.

Employing the ordering shard as a cross-shard coordinator enables processing shards to have the same view on the commitment of cross-shard transactions (i.e., with the same $\mathit{ARGs}$), as each ordering block is public to every processing shard. This can realize a \textit{deterministic commitment}, which is crucial to achieving lock-free execution while guaranteeing cross-shard transactions to be finalized consistently (detailed in Appendix~\ref{appendix-comparison}). Briefly, since each processing shard learns all aggregations in ordering blocks, they can consistently abort failed cross-shard transactions. An initial concern about this approach is that the ordering shard requires extra storage and communication for something irrelevant to transaction ordering. However, we observe that the ordering shard can prune these aggregations once their corresponding cross-shard transactions are finalized since the relevant processing shards provide data availability. Moreover, as we will discuss in \S~\ref{sec-coe-model-optimization} and Appendix~\ref{appendix-extension-details}, we can optimize the data size for aggregations $\mathit{ARGs}$ during the ordering stage. Besides, we also discuss another implementation that frees the ordering shard from the coordination of cross-shard transactions, where processing shards communicate directly to finalize cross-shard transactions, implementing a \textit{leaderless} coordination. However, the leaderless coordination brings more communication overhead as discussed in Appendix~\ref{appendix-extension-details}. 

\textbf{(3) \text{\footnotesize EXECUTE} stage} (Algorithm~\ref{alg-abstract-execution-node-execution}). 
In this stage, processing shards execute and finalize the ordered transactions implied by the ordering block. 
Thanks to the established global order, the process of transaction execution can be simple.
Algorithm~\ref{alg-abstract-execution-node-execution} illustrates a simplified workflow of this stage, and we leave the detailed implementation in Appendix \ref{appendix-detailed-execution-stage} due to the page limitation.

Before executing transactions in a received ordering block $\mathit{OB}$, node $\mathrm{N}_j^{\mathit{sid}}$ in $S_P^{\mathit{sid}}$ traces its locally latest consensus round $\mathit{orderRound}$ and uses a synchronizer to ensure not miss any ordering blocks it involves (lines 1-4). Finalizing transactions consists of three steps (lines 5-9). Abstractly, $\mathrm{N}_j^{\mathit{sid}}$ first finalizes the cross-shard transactions for previous rounds (in $\mathit{ARGs}$) because $\mathit{ARGs}$ aggregates execution results from all relevant processing shards and its corresponding cross-shard transactions are ready for finalization (line 6). Then, $\mathrm{N}_j^{\mathit{sid}}$ executes the ordered intra-shard transactions in $\mathit{\hat{EB}_{dsts}}$ (line 7). Since intra-shard transactions only access the data managed by $S_P^{\mathit{sid}}$, $\mathrm{N}_j^{\mathit{sid}}$ is able to finalize them immediately if they do not involve any uncommitted data (generated by unfinalized cross-shard transactions). Finally, $\mathrm{N}_j^{\mathit{sid}}$ executes the newly ordered cross-shard transactions of $\mathit{\hat{M}_{ctxs}}$ (lines 8-9). Different from intra-shard transactions, the newly cross-shard transactions cannot be finalized in this round as they rely on the execution results from other relevant processing shards. Instead, $\mathrm{N}_j^{\mathit{sid}}$ votes for its local execution results and forward them to its certification module to certify the $\mathit{VOTE}$ result (lines 15-25 in Algorithm \ref{alg-execution-node-certify}).

\begin{algorithm}[t]
    \footnotesize
    \caption{(Simplified) \text{\footnotesize EXECUTE} stage for Node $\mathrm{N}_j^{\mathit{sid}}$ in shard $S^{\mathit{sid}}_P$}
    \label{alg-abstract-execution-node-execution}
    \begin{algorithmic}[1]
        \Statex {\footnotesize \color{Peach}$\blacktriangleright$  execute ordered transactions, $\mathit{OB}= \langle \mathit{ARGs}, \mathit{\hat{EB}_{dsts}}, \mathit{\hat{M}_{ctxs}}, r, \mathit{\sigma^O_{set}}\rangle$}
        \State \textbf{upon} receiving a new ordering block $\mathit{OB}$  \textbf{do}
        \State \hspace*{4mm} \textbf{if} $\mathit{OB.r} > \mathit{orderRound}+1$ and Verify($\mathit{OB}, \mathit{OB.\sigma^O_{set}}$) \textbf{then}
        \State \hspace*{8mm} synchronize missing ordering blocks from $S_{O}$
        \State \hspace*{8mm} update $\mathit{orderRound}$ with the synchronized ordering blocks
        \State \hspace*{4mm} \textbf{elseIf} $\mathit{OB.r} == \mathit{orderRound}+1$ and Verify($\mathit{OB}, \mathit{OB.\sigma^O_{set}}$)
        \Statex{{\footnotesize \color{gray} $\blacktriangleright$ Step 1: handle aggregated results $\mathit{ARGs}$} }
        \State \hspace*{8mm} finalize cross-shard transactions of $\mathit{ARGs}$ for previous rounds
        \Statex{{\footnotesize \color{gray} $\blacktriangleright$ Step 2: handle intra-shard transactions} }
        \State \hspace*{8mm} execute and finalize intra-shard transactions of $\mathit{\hat{EB}_{dsts}}$
        \Statex{{\footnotesize \color{gray} $\blacktriangleright$ Step 3: handle cross-shard transactions} }
        \State \hspace*{8mm} execute and vote for cross-shard transactions of $\mathit{\hat{M}_{ctxs}}$
        \State \hspace*{8mm} generate \text{\scriptsize EXEVOTE} message and send it to the certification module
    \end{algorithmic}
\end{algorithm}

\begin{figure}[ht]
    \centering
    \begin{subfigure}{0.3\textwidth}
        \includegraphics[width=\linewidth]{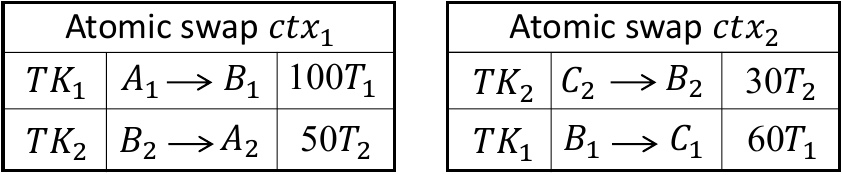}
        \caption{Cross-shard transaction examples.}
        \label{fig:ctx-example}
    \end{subfigure}
    \vspace{1em} 
    
    \begin{subfigure}{0.45\textwidth}
        \includegraphics[width=\linewidth]{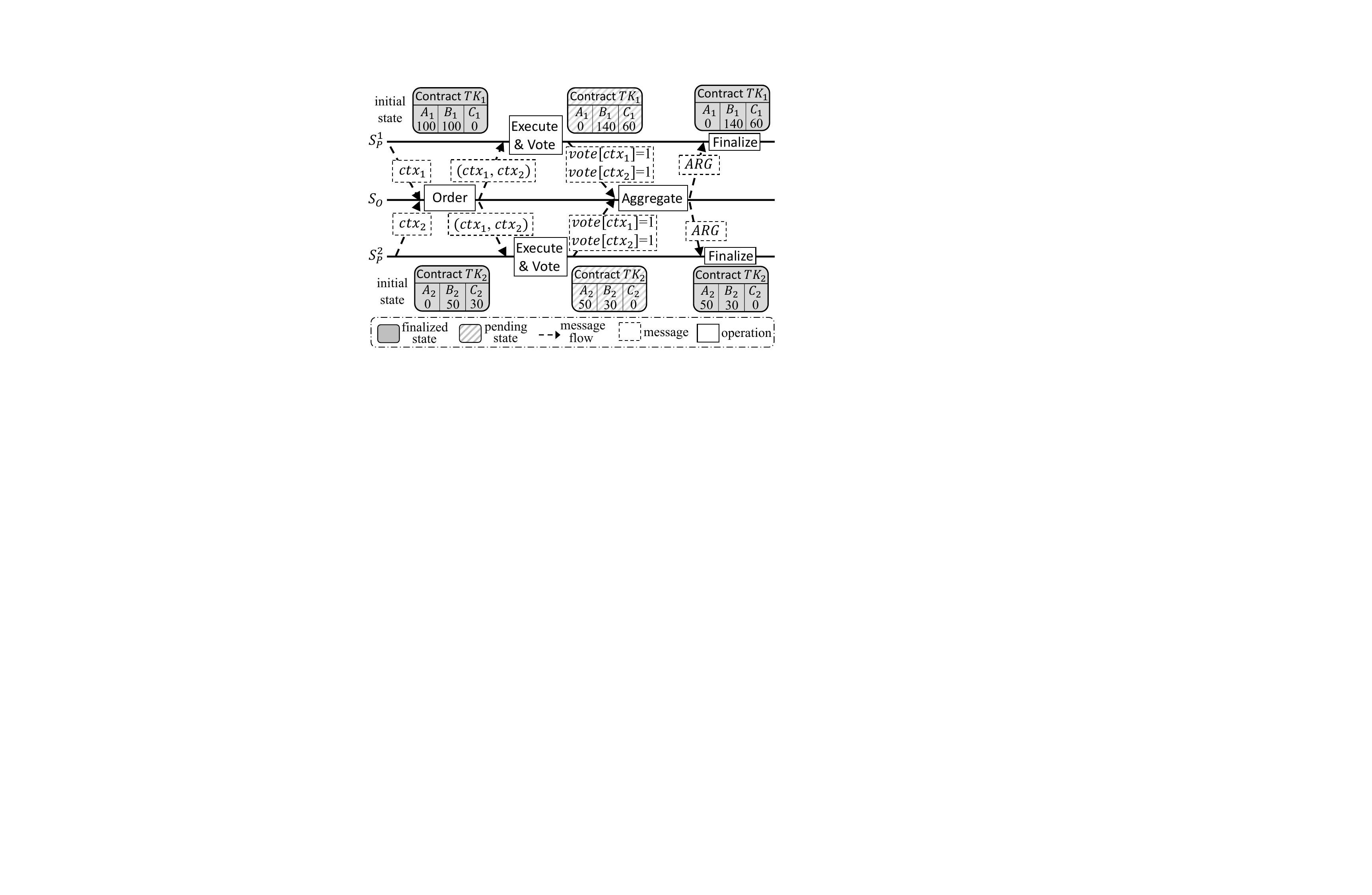}
        \caption{Lock-free execution.}
        \label{fig:lock-free-execution}
    \end{subfigure}
    \caption{The workflow of handling cross-shard transactions in Arete: (a) Two atomic swap transactions $ctx_1$ and $ctx_2$, both are cross-shard and involve contract $TK_1$ managed by shard $S_P^1$ and $TK_2$ managed by $S_P^2$. (b) Arete handles cross-shard transactions $ctx_1$ and $ctx_2$ via a lock-free execution approach.}
    \label{fig:workflow-ctx}
\end{figure}

\noindent \textit{Example:} Arete utilizes the deterministic information provided by the ordering shard to realize a lock-free execution. \Cref{fig:workflow-ctx} illustrates an example that Arete uses the lock-free execution approach to handle two common cross-shard transactions $ctx_1$ and $ctx_2$, both of which involve token swapping between two token smart contracts $TK_1$ (managed by shard $S_P^1$) and $TK_2$ (managed by shard $S_P^2$). Specifically, in $ctx_1$, account $A$ swaps 100$T_1$ with account $B$ for 50$T_2$, and in $ctx_2$, account $B$ swaps 30$T_2$ with account $C$ for 60$T_1$ (\Cref{fig:ctx-example}), where $T_1$ and $T_2$ represent different tokens. First, nodes in $S_P^1$ and $S_P^2$ send $ctx_1$ and $ctx_2$ to the ordering shard $S_O$ via their certificate blocks, respectively. $S_O$ then orders the transactions in an ordering block and forwards the block to $S_P^1$ and $S_P^2$. With the transaction order, $S_P^1$ and $S_P^2$ consistently execute $ctx_1$ and $ctx_2$ without locking relevant states. Each processing shard then sends $\mathit{VOTE}$ messages to $S_O$, indicating whether a cross-shard transaction is executed successfully or not (note that at this time, the modified states are not finalized). $S_O$ aggregates the $\mathit{VOTE}$ messages from $S_P^1$ and $S_P^2$ and sends an aggregator message $\mathit{ARG}$ to them, indicating whether a cross-shard transaction can be committed or not. With $\mathit{ARG}$, $S_P^1$ and $S_P^2$ eventually commit $ctx_1$ and $ctx_2$, ensuring consistency and atomicity. During this process, $ctx_1$ and $ctx_2$ do not lock the states and can be executed within a single consensus instance.

In comparison to previous two-phase commit approaches, the lock-free execution in Arete is more promising in a real-world sharding system where some hotspot smart contracts (e.g., Uniswap~\cite{uniswap}) are frequently accessed. However, a lock-free execution can lead to a \textit{cascading abortion} problem where aborting a cross-shard transaction $ctx$ will make all following transactions conflicting with $ctx$ (i.e., involving in accessing the states modified by $ctx$) aborted. 
While several existing solutions~\cite{guo2021releasing, blockstm} can be integrated into Arete to alleviate the cascading abortion issue, there are also promising new mitigation strategies specifically tailored for our blockchain sharding scenario. For instance, processing shards can avoid certifying new transactions that involve uncommitted states before the corresponding cross-shard transactions are finalized. In fact, the trade-off between aborts (from lock-free execution) and waits (from lock-based execution) has been systematically quantified by previous work~\cite{guo2021releasing}, showing that the performance obtained from the lock-free execution is better than that obtained from the lock-based execution under the scenario with hotspots.

\subsection{Optimizations}\label{sec-coe-model-optimization}
Since Arete employs a single ordering shard to order transactions for all processing shards,
the ordering shard will eventually become the bottleneck as the scale of the network increases. Several optimizations are adopted by our protocol to avoid this bottleneck. 

\textit{Cross-shard transaction batching.} In \S~\ref{sec-coe-specification}, we assume a certificate block contains digests of every cross-shard transaction (i.e., the data field $\mathit{M_{ctxs}}$). However, $\mathit{M_{ctxs}}$ can be replaced by fixed $k-1$ digests in implementation, where $k$ is the number of processing shards in the system. Specifically, a processing shard batches all cross-shard transactions involving the same other processing shard and adds the digest of the batch into the certificate block. Since there are $k-1$ processing shards (excluding itself), a certificate block contains at most $k-1$ digests for cross-shard transaction batches.

\textit{Aggregators compression.} The aggregators $\mathit{ARGs}$ in an ordering block can be compressed as bit strings (see Appendix~\ref{appendix-extension-details} for more details). 
Briefly, since cross-shard transaction batches are explicitly ordered and recorded in previous ordering blocks, the ordering shard can replace the digests of cross-shard transaction batches with a sequence of binary bits, where a bit can represent the finalization result of a cross-shard transaction batch (i.e., 1 means commit while 0 means abort). Since a certificate block contains at most $k-1$ cross-shard transaction batches, $\mathit{ARGs}$ in an ordering block only requires $k-1$ bits for the certificate block.

\textit{Signatures aggregation.} When creating certificate blocks and ordering blocks, we can compress a quorum of signatures (e.g., using BLS~\cite{bls}) to a single signature.

\noindent \textbf{Future bottleneck.} With the above optimizations, we use a similar method as~\cite{danezis2022narwhal} to estimate the future bottleneck of the ordering shard. Specifically, we calculate the ratio of the data volumes that the ordering shard and a processing shard need to handle. Assume we use the SHA-256 algorithm to calculate a digest (i.e., a digest is 32B). As an illustration, an execution block, containing 5,000 transactions of 512B each, a shard ID of 2B, and a node ID of 2B, is around 2.56MB. Its associated certificate block, containing an execution block digest of 32B, $k-1$ batch digests of 32B each, a shard ID of 2B, and an aggregate (BLS) signature of 96B, is $32k+98$B. Eventually, a node in a processing shard needs to handle approximate ($32k$B+$2.56$MB) data to generate a certificate block for the ordering shard. When generating an ordering block, for each certificate block, the ordering shard needs to handle $(k-1)/8$B for its $\mathit{ARGs}$, 32B for its execution block digest $\mathit{EB}_{dst}$, $32(k-1)$B for its digests of cross-shard transaction batches $\mathit{M_{ctxs}}$, $8/k$B for the round number (amortized by the number of processing shards $k$), and $96/k$B for the aggregate signature (amortized by $k$). Eventually, we can calculate the ratio of the data volumes that the ordering shard and a processing shard need to handle by:
\begin{equation}
    \frac{(k-1)/8+32+32(k-1)+8/k+96/k}{32k+2560000}
    \label{eq-bottleneck-ratio}
\end{equation}
Let equation(\ref{eq-bottleneck-ratio})=$\frac{1}{k}$, we get $k\approx 283$, which indicates that we need about 283 processing shards before the ordering shard handles data volume similar to a processing shard. Note that this is a lower bound because it ignores the computation cost of transaction execution required by processing shards, which could be significant once the transaction workload becomes large. Nonetheless, such a lower bound of 283 shards can meet the requirement of a production blockchain network. For instance, the sharding scheme developed by Ethereum (known as Danksharding) plans to split the network into 64 shards to achieve over 100,000 transactions per second~\cite{danksharding}.

\section{Analysis}\label{section-analysis2}
This section proves Arete guarantees sharding safety and sharding liveness. We focus on the security guarantee within each epoch, while the security proof across epochs can be referenced in \textsc{GearBox}~\cite{gearbox}, as our reconfiguration mechanism follows their scheme. We also prove that $\mathcal{P}$-probabilistic liveness in Definition~\ref{def-system-p-probability} indicates $\mathcal{P}$-probabilistic availability in a blockchain sharding system.

\subsection{Sharding Safety Analysis}
Since the ordering shard in Arete runs a BFT consensus to maintain its ledger, the inherent BFT consensus guarantees that nodes in the ordering shard maintain the same prefix ledger. Besides, cross-shard transactions only modify the states of processing shards but not the ordering shard. We therefore only discuss processing shards.
\begin{lemma}\label{lemma-intra-shard-safety}
    For every two honest nodes $N_i^{sid}$ and $N_j^{sid}$ with local ledgers $\mathfrak{L_i}^{sid}$ and $\mathfrak{L_j}^{sid}$ in processing shard $S_P^{sid}$, Arete guarantees: either $\mathfrak{L_i}^{sid} \subseteq \mathfrak{L_j}^{sid}$ or $\mathfrak{L_j}^{sid}\subseteq \mathfrak{L_i}^{sid}$.
\end{lemma}

\begin{myproof}{}
    For the sake of contradiction, assume there is a consensus round $r_c$ where $\mathfrak{L_i}^{sid}$ finalized a list of execution blocks with the digests $\mathit{EB_{dsts}}$, and $\mathfrak{L_j}^{sid}$ finalized another list of execution blocks with the digests $\mathit{EB'_{dsts}}$ ($\mathit{EB_{dsts}} \neq \mathit{EB'_{dsts}}$). However, since $\mathit{EB_{dsts}}$ and $\mathit{EB'_{dsts}}$ are the content of ordering blocks, it means the ordering shard finalizes two ordering blocks in $r_c$, violating the safety of the ordering shard, leading to a contradiction.
\end{myproof}

\begin{lemma}[Cross-shard Atomicity]\label{lemma-inter-shard-finalize}
    For any cross-shard transaction $\mathit{ctx}$ that is finalized by relevant processing shards, all relevant processing shards either commit or abort $\mathit{ctx}$.
\end{lemma}
\begin{myproof}{}
    Without loss of generality, we assume $\mathit{ctx}$ involves two processing shards $S_P^{m}$ and $S_P^{n}$. For the sake of contradiction, assume $S_P^{m}$ commits $\mathit{ctx}$ while $S_P^{n}$ aborts $\mathit{ctx}$. Recall from the COE architecture that the ordering shard picks all aggregators into its ordering blocks. Thus, $S_P^{m}$ receives an aggregator $\mathit{ARG}$ where $\mathit{ARG.vote}[\mathit{M_{ctx}}]=1$ while $S_P^{n}$ receives an aggregator $\mathit{ARG'}$ where $\mathit{ARG'.vote}[\mathit{M_{ctx}}]=0$, or there is another cross-shard transaction $\mathit{ctx'}$ conflicting with $\mathit{ctx}$ such that $\mathit{ARG.vote}[\mathit{M_{ctx'}}]=1$ and $\mathit{ARG'.vote}[\mathit{M_{ctx'}}]=0$. Both of them mean the ordering shard commits two different ordering blocks in the same round. However, the BFT consensus ensures the safety of the ordering shard, leading to a contradiction.
\end{myproof}

\begin{lemma}[Cross-shard Consistency]\label{lemma-inter-shard-order}
    For any two cross-shard transactions $\mathit{ctx_1}$ and $\mathit{ctx_2}$ that are finalized by relevant processing shards, either $\mathit{ctx_1}$ is finalized before $\mathit{ctx_2}$ or $\mathit{ctx_2}$ is finalized before $\mathit{ctx_1}$ in all relevant processing shards.
\end{lemma}
\begin{myproof}{}
    Without loss of generality, we assume $\mathit{ctx_1}$ and $\mathit{ctx_2}$ involve two processing shards $S_P^{m}$ and $S_P^{n}$. For the sake of contradiction, assume $S_P^{m}$ finalizes $\mathit{ctx_1}$ before $\mathit{ctx_2}$ while $S_P^{n}$ finalizes $\mathit{ctx_2}$ before $\mathit{ctx_1}$. 
    Recall from the \text{\footnotesize EXECUTE} stage that cross-shard transactions are finalized in the order of the $\mathit{VOTE}$ results of the aggregators. The above finalization results indicate that $S_P^{m}$ and $S_P^{n}$ receive different aggregators from the ordering shard, which violates the safety of the ordering shard and leads to a contradiction.
\end{myproof}


Lemma \ref{lemma-intra-shard-safety} proves Arete satisfies the condition(i) of Definition~\ref{def-system-safety}, and Lemmas \ref{lemma-inter-shard-finalize} and \ref{lemma-inter-shard-order} prove Arete satisfies the condition(ii) of Definition~\ref{def-system-safety}. Therefore, we have:

\begin{theorem}\label{theorem-system-safety}
    Arete guarantees the sharding safety.
\end{theorem}

\subsection{Sharding Liveness Analysis}

\begin{lemma}\label{lemma-liveness-order}
    Transactions sent to honest nodes are eventually ordered by the ordering shard. 
\end{lemma}
\begin{myproof}{}
    Recall from \S~\ref{subsection-system-architecture} that our shard reconfiguration mechanism ensures at least $f_S\cdot |S_P^{sid}|+1$ nodes in each processing shard $S_P^{sid}$ eventually send their certificate blocks to the ordering shard. Since at most $f_S$ fraction of nodes is Byzantine, the above condition ensures at least one honest node can certify new transactions to the ordering shard. Therefore, every transaction sent to honest nodes can be packed into a certificate block that will be eventually ordered by the ordering shard. 
\end{myproof}

\begin{lemma}\label{lemma-liveness-availability}
    Honest nodes can obtain intact transactions for execution if these transactions have been ordered by the ordering shard.
\end{lemma}
\begin{myproof}{}
    If transactions are ordered by the ordering shard, it means their associated certificate blocks are valid (i.e., containing a quorum of signatures). From Lemma~\ref{lemma-data-availability}, we know that a valid certificate block enables nodes to retrieve all intact transactions corresponding with the block. Therefore, any honest node can eventually obtain these intact transactions.
\end{myproof}

Lemmas \ref{lemma-liveness-order} and \ref{lemma-liveness-availability} jointly ensure each transaction sent to honest nodes can be handled by relevant processing shards and finalized eventually. Therefore, we have:
\begin{theorem}\label{theorem-system-liveness}
    Arete guarantees the sharding liveness.
\end{theorem}

\subsection{Analysis for Probabilistic Liveness}\label{appendix-analysis-p-probability}
It is well-known that a distributed system with crash/Byzantine failures cannot solve the consensus problem deterministically under the asynchronous network (i.e., the FLP impossibility~\cite{fischer1985impossibility}). In other words, a sharding system cannot handle new transactions when the network is asynchronous, i.e., being unavailable or not guaranteeing liveness. Therefore, we only consider that the network is being synchronous in the following proof.
\begin{theorem}\label{theorem-four-nines}
In a sharding system, $\mathcal{P}$-probabilistic liveness indicates $\mathcal{P}$-probabilistic availability.
\end{theorem}
\begin{myproof}{}
    Without loss of generality, we assume the reconfiguration interval is $\mathcal{T}$ years (usually $0<\mathcal{T}<1$), and there are $k$ shards in the system. Recall that $\mathcal{P}$-probabilistic liveness means when forming shards, the probability of a shard that violates liveness is $1-\mathcal{P}$, indicating the expected number of liveness-violated shards is $(1-\mathcal{P})\cdot k$. A liveness-violated shard can be considered unavailable as it neither handles new transactions nor responds to clients. Since liveness-violated shards can be recovered after the shard reconfiguration, $\mathcal{P}$-probabilistic liveness also indicates that $(1-\mathcal{P})\cdot k$ shards in expectation will be unavailable for $\mathcal{T}$ years. When amortizing the unavailable time by all shards, we can calculate the unavailable time of the sharding system during the reconfiguration interval is $\frac{(1-\mathcal{P})\cdot k \cdot \mathcal{T}}{k}=(1-\mathcal{P})\cdot \mathcal{T}$ years. Moreover, since there are $\frac{1}{\mathcal{T}}$ times of shard reconfiguration in a year, the total unavailable time of the sharding system in a year is $(1-\mathcal{P})\cdot \mathcal{T} \cdot \frac{1}{\mathcal{T}}=(1-\mathcal{P})$ years, indicating a $\mathcal{P}$-probabilistic availability of a system.
\end{myproof}

\section{Evaluation}\label{section-evaluation}
\subsection{Implementation}\label{sec-impl}
We implement a prototype for Arete in Rust, which uses Tokio for asynchronous network, ed25519-dalek for elliptic curve-based signature, and RocksDB for persistent storage. The consensus protocol of the ordering shard adopts a variation of Hotstuff~\cite{gelashvili2022jolteon}. For the communications within each shard, we use TCP to realize reliable point-to-point channels. 
For communications between the ordering shard and processing shards, we allow nodes to connect one or more nodes from the other shard randomly. Our implementation involves around 6.5K LOC, and the source codes are public with the testing scripts~\cite{arete-code}.

We compare Arete to two representative sharding protocols: 1) \textsc{GearBox}~\cite{gearbox}, a SOTA sharding protocol focusing on reducing the shard size; 2) RIVET~\cite{rivet}, a sharding scheme that resembles Arete, but uses a leader-based approach for transaction dissemination and does not separate safety and liveness to reduce the shard size (see \Cref{main-body-related-work} for more comparisons). 
Specifically, \textsc{GearBox} separates safety $f_S$ and liveness $f_L$ in each shard, but necessitates $f_S+2f_L < 1$. Each shard runs an intra-shard consensus protocol to commit intra-shard transactions. To handle cross-shard transactions, \textsc{GearBox} adopts a two-phase commit protocol, which locks the involved states of a transaction until all relevant shards finish finalizing it. In \textsc{GearBox}, a specific control shard coordinates the two-phase commit protocol. RIVET deploys a leader node in each processing shard to produce blocks, which are then finalized by a single ordering shard via an intra-shard consensus protocol. Unlike \textsc{GearBox}, RIVET adopts an optimistic cross-shard consensus. Specifically, the ordering shard first orders the cross-shard transactions, and then all relevant processing shards execute them without locking the involved states.
As both \textsc{GearBox} and RIVET do not provide their code implementations, for a fair comparison, we implement and evaluate \textsc{GearBox}
and RIVET
based on our codebase.

\subsection{Experiment Setup}\label{sec-setup}
We run our evaluation in AWS, using c5a.8xlarge EC2 instances spread across 8 regions in the world (3 in Europe, 3 in America, and 2 in Asia). Each instance provides 32 CPUs, 64 GB RAM, and 10 Gbps of bandwidth and runs Linux Ubuntu server 20.04. We deploy one client per node in processing shards to submit transactions at a fixed rate. Each transaction has a size of 512 Bytes. In the following sections, each measurement is the average of 2 runs where shards have been running for 10 minutes to obtain a more precise result.

\begin{figure*}[htbp]
	\centering
        \begin{minipage}[t]{0.24\textwidth}
		\centering
		\includegraphics[width=1.7in]{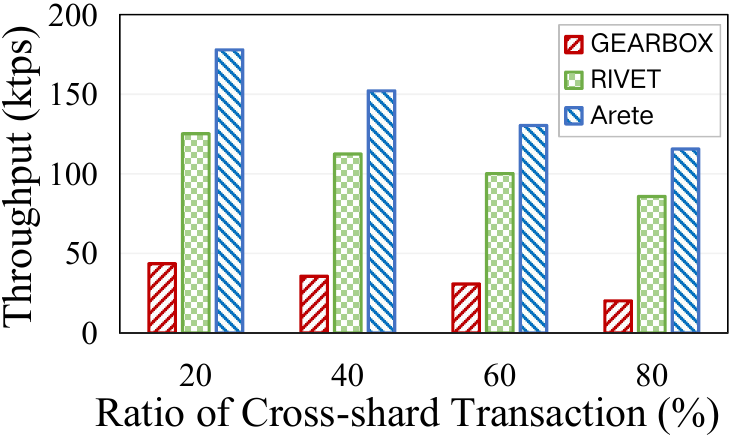}
		\caption{{TPS-CTX ratios}}
		\label{fig-ctx-tps}
	\end{minipage}
        \hfill
        \begin{minipage}[t]{0.24\textwidth}
		\centering
		\includegraphics[width=1.7in]{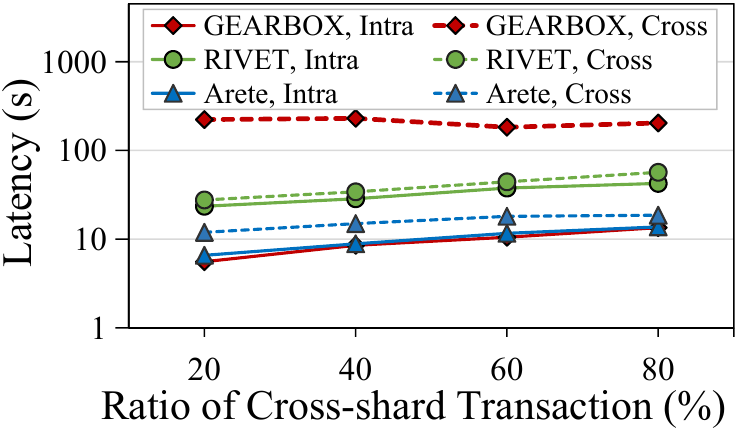}
		\caption{{Latency-CTX ratios}}
		\label{fig-ctx-latency}
	\end{minipage}
        \hfill
	\begin{minipage}[t]{0.24\textwidth}
		\centering
		\includegraphics[width=1.7in]{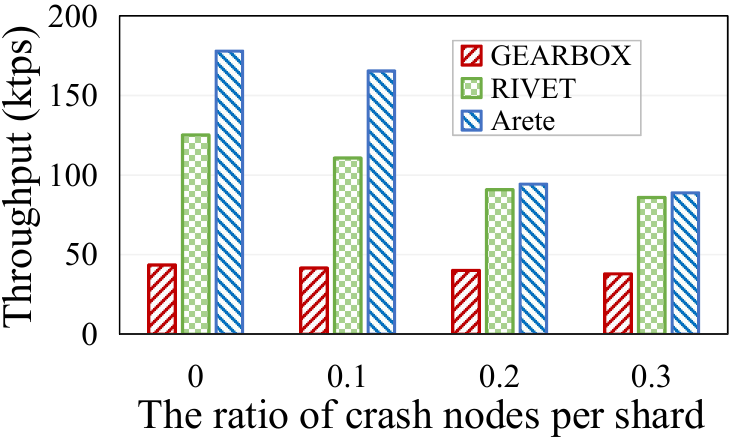}
		\caption{{TPS-Crashes}}
		\label{fig-crash-tps}
	\end{minipage}
        \hfill
	\begin{minipage}[t]{0.24\textwidth}
		\centering
		\includegraphics[width=1.7in]{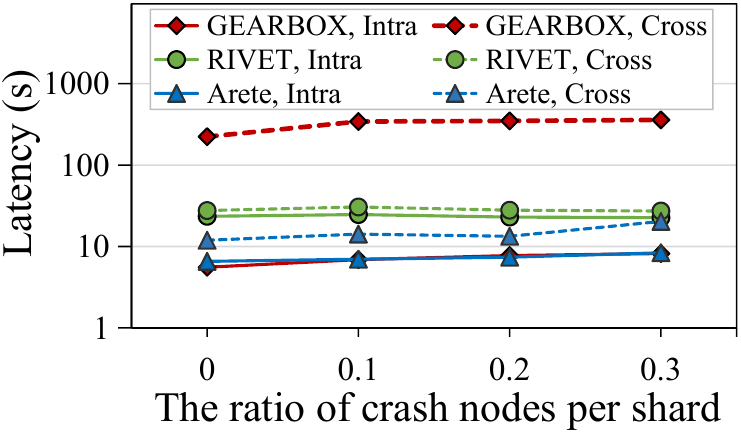}
		\caption{{Latency-Crashes}}
		\label{fig-crash-latency}
	\end{minipage}
\end{figure*}

In the following experiments, we set the total ratio of Byzantine nodes $s=15\%$, the security parameter $\lambda=20$. 
This allows us to create more shards to evaluate the scalability while avoiding excessive monetary expenses on EC2.
For the following measurements, we set the ratio of cross-shard transactions $20\%$ by default but also evaluate its impact on the performance (\Cref{fig-ctx-tps} and \Cref{fig-ctx-latency}). It is worth emphasizing that many solutions~\cite{huang2022brokerchain, zhang2023txallo, optchain, krol2021shard, tao2020sharding} that are proposed to reduce the cross-shard ratio can be applied to Arete. However, we consider them an orthogonal research topic.


\subsection{Scalability}\label{subsection-evaluation-scalability}
We first evaluate the scalability of distinct sharding protocols by running different numbers of nodes $n$. When calculating the shard size, we make sure: (i) the ordering shard (or control shard in \textsc{GearBox}) always satisfies safety and liveness; (ii) the processing shards always satisfy safety, but only guarantee $0.9999$-probabilistic liveness. TABLE~\ref{table-evaluation-parameters} gives the configurations for this experiment, where $m^\#$ represents the size of the ordering (or control) shard, $m^*$ represents the size of the processing shards, $f_L$ represents the liveness threshold of processing shards, and $k$ represents the number of processing shards\footnote{For illustration purposes, we use the term``shard'' to exclusively represent the shard running consensus in \textsc{GearBox} and the processing shard in RIVET and Arete in the following sections, unless explicitly stated otherwise}.

\begin{table}[ht]
    \centering
    \tiny
    \caption{Comparisons under different numbers of nodes ($s=15\%$, $\lambda=20$), shown in the format [\textsc{GearBox}, RIVET, Arete]}
    \label{table-evaluation-parameters}
    \begin{threeparttable}
    \begin{tabular} 
    {c|c c c c c c}
    \hline
    & \multicolumn{6}{c}{The total number of nodes $n$} \\
     & 50 & 100 & 200 & 300 & 400 & 500 \\
    \hline
    $m^\#$ & [21, 21, 21] & [42, 42, 42] & [63, 63, 63] & [72, 72, 72] & [78, 78, 78] & [81, 81, 81]\\
    $m^*$& [20, 15, 13] & [38, 23, 18] & [49, 27, 20] & [57, 29, 22] & [60, 31, 24] & [63, 31, 24]\\
    $f_L(\%)$& [32, 49, 41] & [31, 49, 41] & [31, 49, 42] & [31, 49, 42] & [31, 49, 43] & [31, 49, 43] \\
    $k$& [2, 3, 3] & [2, 4, 5] & [4, 7, 10] & [5, 10, 13] & [6, 12, 16] & [7, 16, 20] \\
    \hline
    \end{tabular}
    \end{threeparttable}
\end{table}


\begin{figure*}[htbp]
	\centering
        \begin{minipage}[t]{0.32\textwidth}
		\centering
		\includegraphics[width=2.3in]{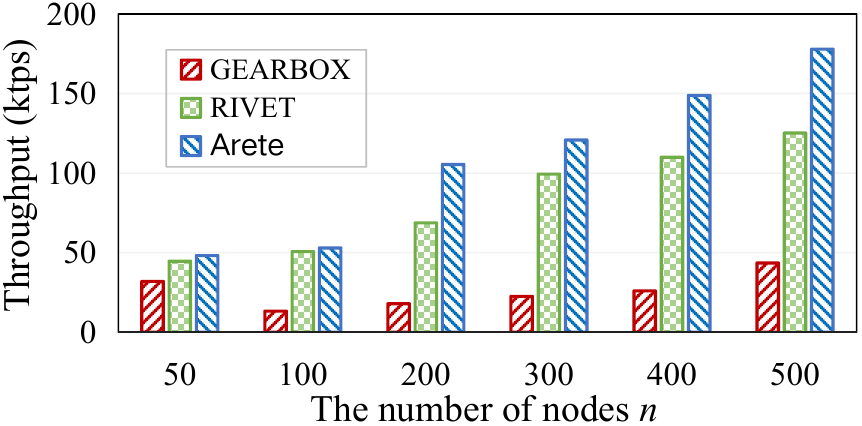}
		\caption{{Throughput-Nodes}}
		\label{fig-nodes-tps}
	\end{minipage}
        \hfill
	\begin{minipage}[t]{0.32\textwidth}
		\centering
		\includegraphics[width=2.3in]{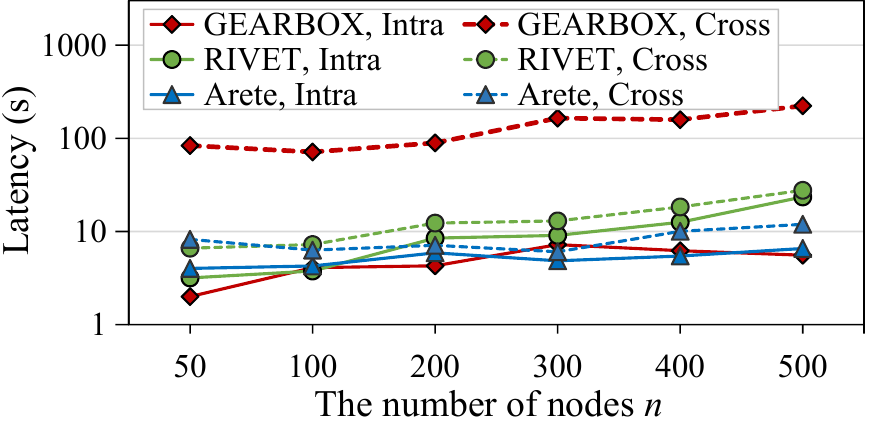}
		\caption{{End-to-end latency-Nodes}}
		\label{fig-nodes-latency}
	\end{minipage}
        \hfill
	\begin{minipage}[t]{0.32\textwidth}
		\centering
		\includegraphics[width=2.3in]{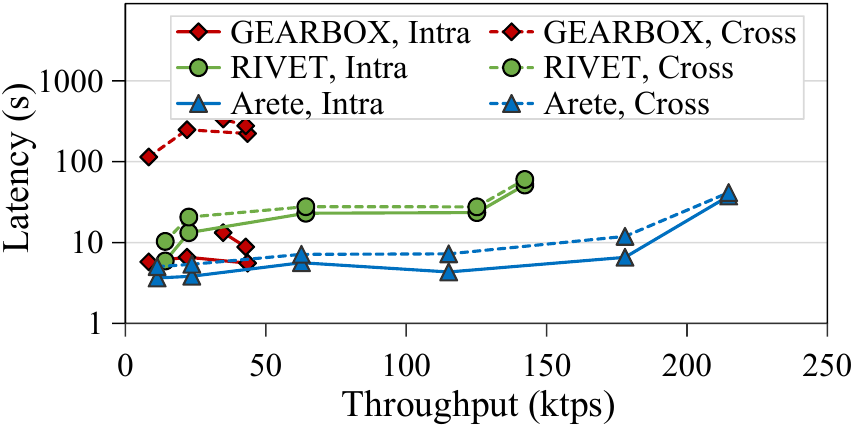}
		\caption{{Throughput-Latency}}
		\label{fig-tps-latency}
	\end{minipage}
\end{figure*}

Figure~\ref{fig-nodes-tps} shows the transaction throughput under different total numbers of nodes and configurations. All compared protocols can scale the system where transaction throughput increases with the increasing number of nodes. There is an exception between $n=50$ and $n=100$, where \textsc{GearBox} achieves higher throughput under $n=50$ than under $n=100$. We find that it is because \textsc{GearBox} has the same number of (i.e., $k=2$) shards but smaller shard sizes when $n=50$ than when $n=100$. 
This shows a larger shard size will compromise transaction throughput as it introduces higher communication overheads. 
Additionally, a smaller shard size enables the system to create more shards and achieve higher throughput. Thus, Arete achieves the best throughput with the same number of nodes, followed by RIVET and then \textsc{GearBox}. This shows that Arete can achieve a better scalability than the compared protocols. Specifically, when $n=500$, Arete has $k=20$ processing shards to handle transactions in parallel and can achieve about 180K transactions per second (TPS), which is 4$\times$ improvement compared to \textsc{GearBox} and 1.4$\times$ improvement compared to RIVET.

Figure~\ref{fig-nodes-latency} shows end-to-end intra-shard and cross-shard confirmation latency under different total numbers of nodes, where we start from the time when the client sends a transaction to the time when the transaction is finalized. When $n$ increases, the confirmation latency of both intra-shard and cross-shard transactions increases since the shard size becomes larger as well. Compared to \textsc{GearBox}, Arete achieves near intra-shard confirmation latency, which varies from approximately 4s to 6s with $n$ increases. For the cross-shard confirmation latency, Arete performs much better than \textsc{GearBox} due to our lock-free execution. To elaborate, the cross-shard confirmation latency of Arete varies from approximately 8s to 12s with $n$ increases whereas \textsc{GearBox} requires 80s to 223s to finalize a cross-shard transaction due to its adopted two-phase commit protocol. Arete reduces the cross-shard confirmation latency by over 10 times compared to \textsc{GearBox}.
Similarly, RIVET achieves lower cross-shard confirmation latency than \textsc{GearBox} thanks to its optimistic cross-shard consensus. However, due to its leader-based block production, RIVET disseminates transactions more slowly and experiences higher confirmation latency than Arete.


\subsection{Performance}
\label{subsection-evaluation-performance}
We evaluate the performance in terms of transaction throughput, end-to-end intra-shard confirmation latency, and end-to-end cross-shard confirmation latency. In the following experiments, we run 500 nodes in total, and the other configurations regarding $m^\#$, $m^*$, $f_L$, and $k$ are illustrated in TABLE~\ref{table-evaluation-parameters} (in the column $n=500$).

\noindent \textbf{Fault-free performance}. We first run the fault-free experiment under different workloads of the system. We use throughput-latency characteristics to depict the performance of the system. The characteristics can evaluate the capacity of a system to handle transactions. To elaborate, before the workload reaches saturated, the confirmation latency changes slightly while transaction throughput can increase noticeably as the workload enlarges; in contrast, after the workload is saturated, transaction throughput will become steady while the confirmation latency will increase noticeably as the workload enlarges. Therefore, the steady transaction throughput is used to evaluate the optimal performance of a system. Figure~\ref{fig-tps-latency} shows the throughput-latency characteristics of compared protocols. Under $n=500$, Arete can achieve an optimal performance of about 180K transaction throughput at an intra-shard latency below 7 seconds and a cross-shard latency below 12 seconds. In contrast, the optimal transaction throughput of \textsc{GearBox} is around 45K while its cross-shard latency is much higher than its intra-shard latency. 
Furthermore, the optimal transaction throughput of RIVET is around 125K at an intra-shard latency of around 23 seconds and a cross-shard latency of around 27 seconds.
Intuitively, the high throughput of Arete benefits from smaller shard sizes and more shards. Furthermore, a larger liveness threshold $f_L$ also enables Arete to move more quickly to the next phase since the quorum of moving to the next phase (1-$f_L$) decreases, and a leaderless approach amortizes the data dissemination overhead. This validates the benefits of reducing shard sizes and enhancing the transaction process with the proposed COE architecture.

\noindent \textbf{Performance under cross-shard transaction ratios}. We then compare the performance of distinct protocols under different ratios of cross-shard transactions (CTXs). Specifically, we set the cross-shard ratio to $20\%$, $40\%$, $60\%$, and $80\%$. A higher cross-shard ratio introduces more overheads to the system as the ordering shard (or the control shard in \textsc{GearBox}) needs to handle more data relevant to CTXs. \Cref{fig-ctx-tps} and \Cref{fig-ctx-latency} respectively show the throughput and latency under varying cross-shard ratios. We find that as the cross-shard ratio increases, the throughput of all protocols decreases, and their confirmation latency increases. However, Arete outperforms both \textsc{GearBox} and RIVET regardless of the cross-shard ratios.

\noindent \textbf{Performance under crash faults}. We finally evaluate the performance under the crash fault model. We conduct experiments with $0\%, 10\%, 20\%$, and $30\%$ ratios of crash nodes in each shard respectively. A crash node in our experiments neither responds to clients nor participates in the sharding protocol. Figure~\ref{fig-crash-tps} shows the transaction throughput under different crashed ratios. 
We observe that the number of crash nodes can compromise the throughput of Arete more than the throughput of \textsc{GearBox} and RIVET. Specifically, the throughput of Arete drops from 180K to 88K as the ratio of crash nodes increases. In contrast, the throughput of \textsc{GearBox} drops slightly (approximately 45K to 38K) as the ratio of crash nodes increases; the throughput of RIVET drops approximately from 125K to 86K.
This indicates that nodes in Arete can contribute their bandwidth to the transaction throughput more effectively than both \textsc{GearBox} and RIVET as we deploy a leaderless approach to disseminate transactions. 
Figure~\ref{fig-crash-latency} shows the latency of protocols under different crashed ratios. We observe that with more crash faults in the system, both intra-shard and cross-shard confirmation latencies of all protocols increase. Intuitively, due to fewer active nodes in each shard, a shard requires a larger ratio of nodes' signatures to move to the next stage. In this case, any slight network delay between two nodes may significantly slow down the process of finalizing transactions.

\section{Related Work}
\label{main-body-related-work}
\noindent\textbf{Blockchain sharding}.
Sharding protocols~\cite{Elastico, COSPLIT, redbelly, chainspace, Omniledger, Rapidchain, Monoxide, ahl, rivet, xutwo, sharper, byshard, amiri2019sharding, hong2023prophet} are proposed to scale blockchains. 
While a recent line of orthogonal sharding protocols~\cite{huang2022brokerchain, zhang2023txallo, hong2023prophet, liu2024kronos, jiang2024sharon} focuses on handling cross-shard transactions, this paper focuses on resolving the size-security dilemma. For this topic, many works~\cite{rivet, instachain, gearbox, licochain, xutwo} are designed to create smaller shards in a larger quantity. CoChain~\cite{licochain} and DL-Chain~\cite{lin2024dl} allow the creation of a large number of small corrupt shards and propose a Consensus-on-Consensus (CoC) framework to finalize consensus results. Nevertheless, CoC introduces extra overheads, as handling transactions requires one more consensus instance, and each shard needs to serve multiple CoC committees. Reticulum~\cite{xutwo} proposes a two-layer sharding protocol, where multiple process shards that propose blocks form a control shard to finalize the proposed blocks. By separating safety and liveness in process shards and resorting to the corresponding control shard for block finalization, Reticulum allows for more lightweight process shards to be created and enhances the scalability. We highlight several key differences between Reticulum and Arete. First, Reticulum relies on the synchrony network assumption for its security, while Arete operates in the partial synchrony network. Second, the process shard in Reticulum adopts a leader-based scheme to propose blocks, where only one block is proposed each time, and the leader could potentially become the bottleneck; in contrast, the processing shard in Arete adopts a leaderless scheme to produce blocks, fully utilizing nodes' resources. Last but not least, Arete employs a single ordering shard to establish a global order, facilitating the finalization of cross-shard transactions, while cross-shard transaction handling cannot benefit from Reticulum's design.

RIVET~\cite{rivet} proposed a reference-worker sharding scheme that assigns a single reference shard for ordering, resembling our scheme but with notable distinctions. In Arete, we decouple and pipeline all SMR tasks, allowing shards to perform these tasks asynchronously and in parallel. In contrast, RIVET exclusively decouples the ordering task from the data dissemination and execution, where data dissemination and execution remain coupled and sequential. Moreover, RIVET employs an execute-order-commit architecture to handle intra-shard transactions, which not only necessitates a leader in each worker shard to coordinate the execution stage, but could also lead to a high transaction abortion due to potentially inconsistent order. In contrast, our processing shards perform in a leaderless way, and the global order established by the ordering shard can avoid the order inconsistency issue.


\noindent\textbf{Order-execution decoupling}. Many blockchain protocols decouple execution from ordering to enhance performance. 
Hyperledger~\cite{hyperledger} proposes an execute-order-validate (EOV) architecture to support paralleling transaction execution before ordering. Nathan et al.~\cite{fabricssi} and BIDL~\cite{bidl} adopt an execute-order-parallel-validate that can hide the ordering cost by decoupling and parallelizing the transaction execution and ordering. However, they can incur read/write conflicts due to inconsistent access, leading to transaction abortion and compromising performance. Many works~\cite{sharma2019blurring, ruan2020transactional} strive to reduce the abortion rate by applying some reordering algorithms but still suffer from a high abortion rate when the transaction workload involves frequent conflicts. In contrast, ACE~\cite{ACE}, SaberLedger~\cite{liu2021parallel}, and ParBlockchain~\cite{amiri2019parblockchain} adopt an order-execute (OE) architecture and support parallelized execution. Many other works adopt a consensus/ordering-free decoupling architecture for transaction execution. These solutions can eliminate the transaction abortion brought by the EOV architecture. However, they either rely on a single trusted server (e.g., Neuchain~\cite{neuchain}) or solely support a simple transfer application~\cite{fastpay, cryptoconcurrency, orderlesschain}.
Moreover, all the above (partially) decoupling architectures still couple the data dissemination of SMR with ordering/execution, which has been pointed out as a main bottleneck in a high-performance SMR~\cite{prism, ohie, giridharan2024motorway, danezis2022narwhal}. 



\noindent\textbf{Disperse-order-execution decoupling}. Recent works Star~\cite{duan2024dashing} and Motorway~\cite{giridharan2024motorway} show an incredible performance by fully decoupling SMR. However, they are designed for non-sharding blockchains and have limited scalability as a large network brings heavy communication overheads to nodes. Pando~\cite{wang2024pando} proposes a scalable decoupling SMR by optimizing communication with a sampling technique but fails to support paralleled execution compared to sharding protocols. The DAG-based BFT protocols~\cite{keidar2021all, danezis2022narwhal, spiegelman2022bullshark, shoal, shoal++, suilutris, sailfish, babel2023mysticeti} also fully separating three SMR tasks. They adopt a vertical scaling architecture where newly joining nodes (called workers) must trust the primary node they associate with, introducing a stronger trust assumption. In contrast, Arete adopts a horizontal sharding architecture without bringing new trust assumptions. Furthermore, many DAG-based BFT protocols~\cite{danezis2022narwhal, spiegelman2022bullshark} depend on reliable broadcast to disseminate transactions/blocks in the \text{\footnotesize CERTIFY} stage, as they rely on the creation of non-equivocated blocks to complete the ordering task. In contrast, Arete, like~\cite{duan2024dashing, giridharan2024motorway}, is free from reliable broadcast as its ordering task can be completed by the ordering shard independently of block creation.

\section{Conclusion}
\label{section-conclusion}

This work proposed Arete, an optimal sharding protocol achieving scalable blockchains with deconstructed SMR. 
The extensive experiments running on the AWS environment show that Arete outperforms the state-of-the-art sharding protocols in terms of scalability, throughput, and latency.





\section*{Acknowledgment}
\noindent We thank our anonymous reviewers for helpful suggestions about the paper. We also thank Ling Ren and Sourav Das for their insightful discussion about their RIVET project. This work was supported in part by NIFA award number 2021-67021-34252, the National Science Foundation (NSF) under grant CNS1846316. Any opinions, findings, conclusions, or recommendations expressed in this material are those of the authors and do not necessarily reflect the views of the sponsors.

\balance

\bibliographystyle{unsrt}
\bibliography{cite}

\appendices
\section{Proofs of Fault Tolerances in SMR}
\label{appendix-proof-tolerance}
In this section, we prove the Byzantine fault tolerance of each SMR task with the safety-liveness separation. Similar to the aforementioned notations, let $f$ denote the security threshold, $f_S$ denote the safety threshold, and $f_L$ denote the liveness threshold. A Byzantine node is called a safety fault if it attempts to break the safety property by, for example, proposing two different blocks in the same position of the blockchain ledger. Similarly, a liveness fault is defined as a node attempting to break the liveness property by, for example, refusing to respond to honest nodes. With the definitions, there are up to an $f_S$ fraction of safety faults and an $f_L$ fraction of liveness faults in the SMR. Moreover, we consider a partially synchronous network model as assumed in the paper, where message delays are bounded by $\Delta$ after GST. Recall that the safety and liveness of the SMR are implicated if all SMR tasks can be completed and their goals can be achieved. Therefore, the key idea of our proofs is to show each SMR task can be completed, and its goal can be achieved with given $f_S$ and $f_L$.

For the data dissemination task, recall that it is completed if new transactions are dispersed and its goal is to ensure all honest nodes can retrieve an intact block if the transactions are ordered and committed.

\begin{lemma}\label{lemma-dissemination-threshold}
    The data dissemination task of SMR can disperse new transactions such that all honest nodes can retrieve the intact transactions after they are ordered and committed, as long as $f_S+f_L<1$.
\end{lemma}
\begin{myproof}{}
    Since there is at least one honest node in SMR, at least one block of transactions $B$ can be generated by honest nodes. We know that as long as one honest node receives this intact block, all honest nodes will be able to retrieve it by synchronizing it from the honest node. To this end, the honest creator of $B$ disseminates $B$ to the network and collects signatures on it from other nodes. Since it is impossible to distinguish a liveness fault (that does not send messages) and a slow honest node whose messages are in the flight due to the network delay, the honest creator has to move on after receiving a quorum $(1-f_L)$ signatures indicating that the other nodes have received the intact $B$. Note that the process of collecting a quorum $(1-f_L)$ signatures can be terminated after GST. Because of $f_S+f_L<1$, the quorum $(1-f_L)$ ensures at least one honest node receives the intact $B$. In this case, all honest nodes can eventually retrieve the intact $B$ by synchronizing it from the honest node storing $B$, if $B$ is later ordered and committed. In other words, the data dissemination task can be completed with its goal achieved.
\end{myproof}

For the ordering task, recall that its goal is to ensure all honest nodes output blocks (of transactions) in the same order. One can run an instance of consensus to implement this task. Therefore, the security proof of a partially synchronous consensus protocol can be directly referred to. To elaborate, we have the following lemma:
\begin{lemma}\label{lemma-ordering-threshold}
    The ordering task of SMR can order and output blocks into a tamper-proof ledger such that all honest nodes have the same order for blocks recorded in their local ledger, as long as $f_S+2f_L<1$.
\end{lemma}
\begin{myproof}{}
    For the sake of contradiction, assume honest nodes output two orders $\xi$ and $\xi'$ after completing the ordering task, where $\xi \neq \xi'$. Similarly, due to the indistinguishability of liveness faults and slow honest nodes, a node has to finish the ordering task after receiving a quorum $(1-f_L)$ signed confirmations/votes from other nodes. Let $C$ and $C'$ be the two sets of $(1-f_L)$ quorum of nodes that have outputted $\xi$ and $\xi'$. As $C$ and $C'$ intersect at $2(1-f_L)-1 > f_S$ quorum of nodes (remember we have the condition $f_S+2f_L<1$), at least an honest node must output for both $\xi$ and $\xi'$, leading a contradiction. Moreover, because message delays are bounded after GST, honest nodes can eventually terminate with the received $(1-f_L)$ messages.
\end{myproof}

For the execution task, recall that it must ensure all honest nodes have the same and correct execution results which are externally verified. As the blockchain system adopts a deterministic execution engine, namely, each node generates the same execution results with the same initial state and inputs (i.e., the ordered blocks from the ordering task). Since the ordering task guarantees every honest node has the same order of blocks (i.e., without equivocations), all honest nodes can output the same execution results. As a result, we only need to prove the correctness of execution results are externally verified. Specifically, an external party (e.g., a client) can verify the execution results without participating in the execution task (a.k.a., external verifiability~\cite{momose2021multi, das2022spurt}). To elaborate, we have the following lemma:
\begin{lemma}\label{lemma-execution-threshold}
    The execution task can generate execution results whose correctness can be verified by an external party, as long as $f_S+f_L<1$.
\end{lemma}
\begin{myproof}{}
    Similarly, due to the indistinguishability of liveness faults and slow honest nodes, an external party can only collect a quorum $(1-f_L)$ outputs from nodes before it can verify the correctness of execution results. Note that such a process of collecting a quorum $(1-f_L)$ outputs can be terminated after GST. Because of $f_S+f_L<1$, the $1-f_L > f_S$ quorum contains at least one output from honest nodes. As a result, the external party learns that the execution results are correct.
\end{myproof}

\noindent \textbf{Fault tolerances without safety-liveness separation.} From Lemma \ref{lemma-dissemination-threshold}, \ref{lemma-ordering-threshold}, and \ref{lemma-execution-threshold}, we can conclude the security fault tolerance of each task of the SMR without the safety-liveness separation, by making $f_S=f_L$. To elaborate, the data dissemination and execution tasks can tolerate up to $f<1/2$ Byzantine nodes, while the ordering task can tolerate only $f<1/3$ Byzantine nodes.

\begin{figure*}
    \centering
    \includegraphics[width=7.3in]{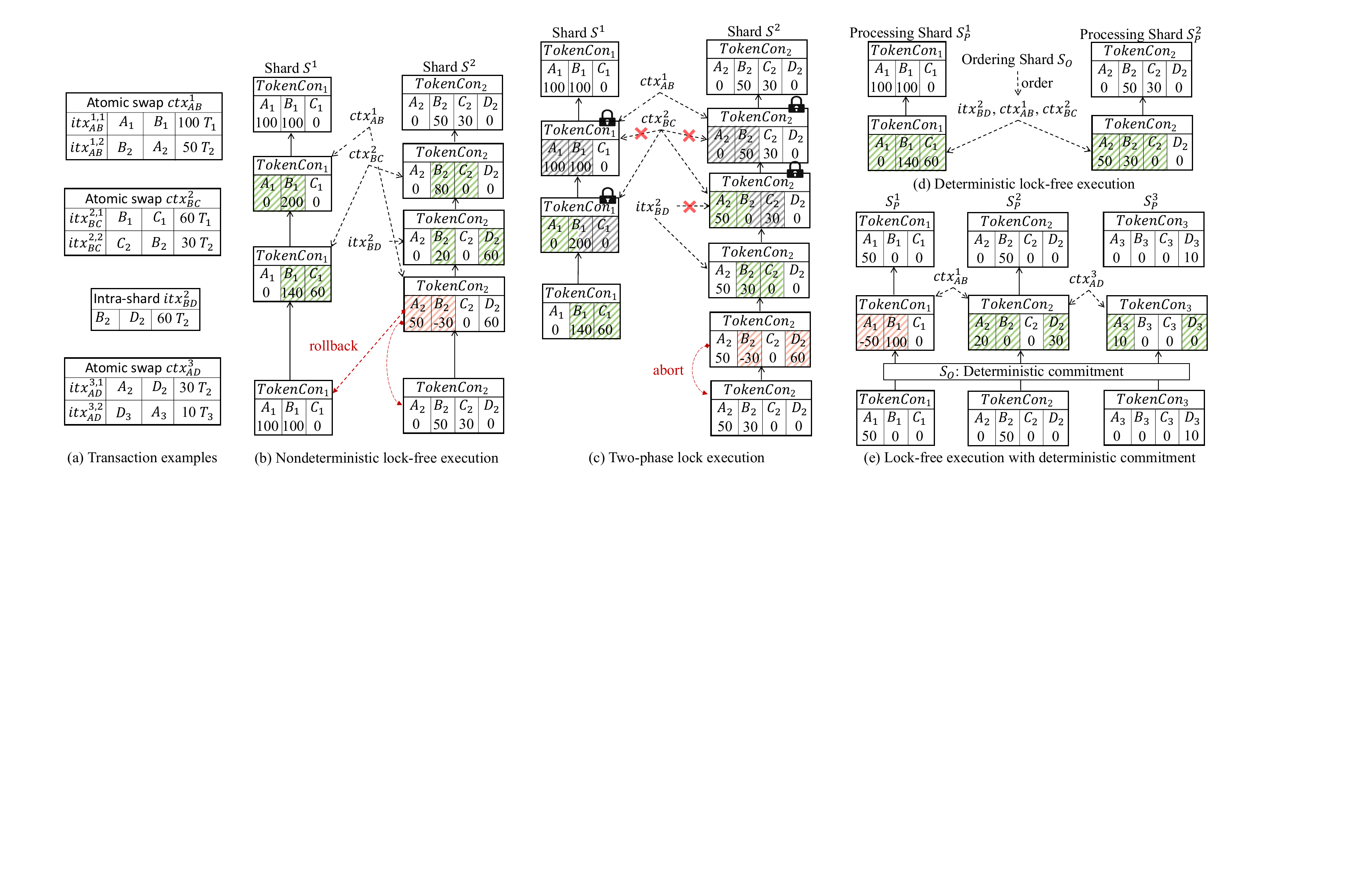}
    \caption{Comparison between different cross-shard execution approaches: (a) illustrative transactions invoking $TokenCon$ contracts to exchange different tokens $T_1$-$T_3$; (b) Optimistic execution without a deterministic order; (c) two-phase lock execution; (d) and (e) A deterministic order achieving lock-free execution. Successfully committing states are highlighted with green color; unsuccessfully modified states are highlighted with red color; locking states are highlighted with gray color.}
    \label{fig-comp-cross-execution}
\end{figure*}
\section{Comparison of Cross-shard Approaches}
\label{appendix-comparison}
While enabling multiple shards to process transactions in parallel, a blockchain sharding protocol introduces concurrency problems as well. Specifically, since cross-shard transactions involve multiple states maintained by multiple shards, there exists \textit{race conditions} where cross-shard and intra-shard transactions can concurrently access and modify the same states. The sharding system must prevent different shards from generating inconsistent reads/writes on the same states. To this end, various cross-shard execution approaches are proposed to eliminate race conditions in blockchain sharding systems. In this section, we will compare our lock-free execution approach with other cross-shard execution approaches adopted by previous sharding systems.

\noindent \textbf{Illustrative transaction examples}. For illustration purposes, we assume contracts $\mathit{TokenCon_1}$, $\mathit{TokenCon_2}$, and $\mathit{TokenCon_3}$ are deployed in different shards $S^1$, $S^2$, and $S^3$. Each $\mathit{TokenCon}$ contract maintains balances of the token for all their users. Different $\mathit{TokenCon}$ contract has its own token (denoted by $T_1$, $T_2$, and $T_3$, respectively), and different tokens can be exchanged with each other at various prices\footnote{For instance, in Ethereum, users can trade different tokens (such as SAI, SHIB, and LINK) from decentralized exchanges (such as Uniswap).}. The blockchain sharding system supports token exchanges across multiple shards via a so-called \textit{atomic swap transaction}. An atomic swap transaction is a cross-shard transaction that involves two $\mathit{TokenCon}$ contracts deployed by two different shards. The result of processing an atomic swap transaction, for example, is a user $A$ transfers token $T_1$ to another user $B$ in $\mathit{TokenCon_1}$ while $A$ obtains token $T_2$ from $B$ in $TokenCon_2$. To achieve this, an atomic swap transaction will be divided into two intra-shard transactions that are sent to relevant shards to complete the transfer operations atomically.

Figure~\ref{fig-comp-cross-execution}(a) gives several examples of transactions that will be used to illustrate different cross-shard execution approaches. In particular, $\mathit{ctx^1_{AB}}$ is an atomic swap transaction where user $A$ trades 100 $T_1$ for 50 $T_2$ with user $B$. $\mathit{ctx^2_{BC}}$ is an atomic swap transaction where user $B$ trades 60 $T_1$ for 30 $T_2$ with user $C$. $\mathit{ctx^3_{AD}}$ is an atomic swap transaction where user $A$ trades 30 $T_2$ for 10 $T_3$ with user $D$. There is an intra-shard transaction $\mathit{itx^2_{BD}}$ that calls the contract $\mathit{TokenCon_2}$ to transfer 60 $T_2$ from user $B$ to user $C$. Recall that processing an atomic swap transaction will create two intra-shard transactions. For instance, $\mathit{ctx^1_{AB}}$ consists of two intra-shard transactions $\mathit{itx^{1,1}_{AB}}$ and $\mathit{itx^{1,2}_{AB}}$. In shard $S^1$, $\mathit{itx^{1,1}_{AB}}$ will invoke $\mathit{TokenCon_1}$ contract to transfer 100 $T_1$ from $A$'s account (denoted by $A_1$) to $B$'s account (denoted by $B_1$). Similarly, in shard $S^2$, $\mathit{itx^{1,2}_{AB}}$ will invoke $\mathit{TokenCon_2}$ contract to transfer 50 $T_2$ from $B$'s account (denoted by $B_2$) to $A$'s account (denoted by $A_2$). We now demonstrate that different cross-shard execution approaches process these illustrative transactions in the following.

\noindent \textbf{Optimistic execution without a deterministic order}. Figure~\ref{fig-comp-cross-execution}(b) illustrates an optimistic execution approach for cross-shard transactions, which is adopted by some previous sharding protocols~\cite{Monoxide, rivet}. Specifically, the optimistic execution approach handles cross-shard transactions in an optimistic way: each shard first independently executes and commits intra-shard transactions that are created by atomic swap transactions; then if there are some inconsistent reads/writes or failed executions, all involved transactions (both intra-shard and cross-shard) require rollback.

The above optimistic approach achieves a lock-free execution for cross-shard transactions. However, it can lead to an unpredictable rollback, compromising the performance and making it impractical in some scenarios where the on-chain rollbacks can bring irreversible off-chain losses (e.g., using cryptocurrencies to buy food). In the example shown in Figure~\ref{fig-comp-cross-execution}(b), $S^1$ modifies $\mathit{TokenCon_1}$ by serially executing $\mathit{itx_{AB}^{1,1}}$ and $\mathit{itx_{BC}^{2,1}}$; $S^2$ modifies $\mathit{TokenCon_2}$ by serially executing $\mathit{itx_{BC}^{2,2}}$, $\mathit{itx^2_{BD}}$, and $\mathit{itx_{AB}^{1,2}}$. In this case, $S^1$ and $S^2$ execute their relevant cross-shard transactions $\mathit{ctx^1_{AB}}$ and $\mathit{ctx^2_{BC}}$ in different orders. A rollback happens when $S^2$ detects $\mathit{itx_{AB}^{1,2}}$ is failed to be executed (due to the race condition from $\mathit{itx^2_{BD}}$). As a result, both $S^1$ and $S^2$ are required to discard all modified states generated by $\mathit{ctx_{AB}^1}$, $\mathit{ctx_{BC}^2}$, and $\mathit{itx^2_{BD}}$. To conclude, inconsistent execution orders lead to inconsistent reads/writes in the same states, which leads to the failure of a transaction. In the optimistic approach, the failure of a transaction can eventually lead to rollbacks of multiple transactions, compromising performance. 

\noindent \textbf{Two-phase lock-based commit}. Most previous sharding protocols, instead, adopt another cross-shard execution approach - the two-phase commit approach. The two-phase commit approach is driven by a coordinator (e.g., a client~\cite{androulaki2018channels, Omniledger} or a shard~\cite{chainspace, Rapidchain, ahl, byshard, zhang2023front}) consisting of two phases: (i) voting phase where the coordinator sends the divided intra-shard transactions to all relevant shards to execute and lock involved states, and (ii) committing phase where the coordinator collects the vote results and informs all relevant shards either commit (if all vote for commit) or abort (if otherwise) the transaction. When states are locked by a cross-shard transaction during the voting phase, any other intra-shard or cross-shard transactions cannot access the locked states and will be blocked from being executed until the cross-shard transaction finishes its committing phase and releases the locks.

Figure~\ref{fig-comp-cross-execution}(c) shows an example of the two-phase commit/lock approach. From top to bottom, $S^1$ and $S^2$ first lock and execute $\mathit{ctx_{AB}^1}$. Before $\mathit{ctx_{AB}^1}$ is committed, $\mathit{ctx_{BC}^2}$ is refused to invoke the relevant contracts $\mathit{TokenCon_1}$ and $\mathit{TokenCon_2}$ that are locked by $\mathit{ctx_{AB}^1}$. Similarly, the transaction $\mathit{itx^2_{BD}}$ is blocked from being executed until $\mathit{ctx_{BC}^2}$ is committed and releases its lock on $\mathit{TokenCon_2}$. It can be seen that the two-phase lock execution for cross-shard transactions prevents shards from accessing states inconsistently and thus eliminates race conditions. The main drawback of the two-phase lock approach is its long confirmation latency. To elaborate, both intra-shard and cross-shard transactions have to wait for a long locking time before they can be executed. This drawback is also shown in our experiments where the two-phase commit approach has $20\times$ cross-shard confirmation latency compared with our protocol. Such a long confirmation latency is unacceptable in a real-world scenario where there always exist some popular smart contracts that are invoked frequently, e.g., the decentralized exchange market Uniswap.

\noindent \textbf{Deterministic lock-free execution}. Different from previous cross-shard execution approaches, this paper proposes another solution built on the global order established by our ordering shard, as shown in Figure~\ref{fig-comp-cross-execution}(d). The deterministic lock-free execution adopted by us addresses the limitations of the two previous cross-shard execution approaches in two aspects. On the one hand, the ordering shard in our protocol establishes a deterministic global order, which prevents failed executions introduced by inconsistent reads/writes that exist in previous optimistic execution approaches. On the other hand, a deterministic global order allows all processing shards to access the states consistently (i.e., each processing shard has a consistent serial order for transactions), by which states are no longer locked to prevent race conditions, and thus transactions can be executed in a lock-free way.

However, a deterministic order is insufficient to achieve a consistent commitment, i.e., all processing shards either commit or abort all their relevant transactions. The inconsistent commitment comes from some \textit{implicit dependencies} of cross-shard transactions across multiple shards. To elaborate, we assume there are three processing shards $S_P^1$, $S_P^2$, and $S_P^3$ processing cross-shard transactions $\mathit{ctx_{AB}^1}$ and $\mathit{ctx_{AD}^3}$ where $\mathit{ctx_{AB}^1}$ is ordered before $\mathit{ctx_{AD}^3}$. As illustrated in Figure~\ref{fig-comp-cross-execution}(e), $S_P^1$ fails to execute $\mathit{ctx_{AB}^1}$ as $A$'s account in $\mathit{TokenCon_1}$ has no enough $T_1$ tokens, and $\mathit{ctx_{AB}^1}$ is eventually aborted by $S_P^1$ and $S_P^2$. However, due to the lock-free execution, $\mathit{ctx_{AD}^3}$ accesses the modified states generated by $\mathit{itx_{AB}^{1,2}}$ of $\mathit{ctx_{AB}^1}$, and $S_P^2$ will vote to commit $\mathit{ctx_{AD}^3}$ (at the time $S_P^2$ does not know $\mathit{ctx_{AB}^1}$ will be aborted since $\mathit{itx_{AB}^{1,2}}$ is executed successfully from its view). Recall from Algorithm~\ref{alg-ordering-node-order} that the aggregated vote on $\mathit{ctx_{AD}^3}$ will be set true (i.e., 1) because the ordering shard receives the $\mathit{VOTE}$ results that indicate $\mathit{ctx_{AD}^3}$ is executed successfully by both $S_P^2$ and $S_P^3$. This can lead to inconsistent commitment where $S_P^2$ aborts $\mathit{ctx_{AD}^3}$ (due to $\mathit{ctx_{AD}^3}$ accesses the aborted states of $\mathit{ctx_{AB}^1}$), while $S_P^3$ commits $\mathit{ctx_{AD}^3}$ (since $S_P^3$ learns the true aggregated vote for $\mathit{ctx_{AD}^3}$ from the ordering shard). 

Arete solves this problem with a \textit{deterministic commitment} mechanism, as shown in Figure~\ref{fig-comp-cross-execution}(e). Specifically, all aggregations of cross-shard transactions are piggybacked on ordering blocks, by which all processing shards know which cross-shard transactions are aborted and can consistently abort the cross-shard transactions based on the dependencies of their states. For instance, with the aggregations, $S_P^3$ knows that $\mathit{ctx_{AB}^1}$ is aborted and $\mathit{ctx_{AB}^1}$ involves $A$'s account in $\mathit{TokenCon_2}$. Since $\mathit{ctx_{AD}^3}$ also involves $A$'s account in $\mathit{TokenCon_2}$, $S_P^3$ will eventually abort $\mathit{ctx_{AD}^3}$. 

\begin{algorithm}[!b]
    \footnotesize
    \caption{\text{\footnotesize EXECUTE} stage for Node $\mathrm{N}_j^{\mathit{sid}}$ in shard $S^{\mathit{sid}}_P$}
    \label{alg-execution-node-execution}
    \begin{algorithmic}[1]
        \Statex \textbf{Local data}: $\mathfrak{L}^{\mathit{sid}}$, $\mathcal{M}$, $\mathit{orderRound}$, $\mathit{curPending}$, $\mathit{curAborted}$, $\mathit{orderedItxs}$ 
        \Statex {\footnotesize \color{Peach}$\blacktriangleright$  execute the ordered transactions}
        \State \textbf{upon} receiving a new ordering block $\mathit{OB}$  \textbf{do}
        \State \hspace*{4mm} \textbf{if} $\mathit{OB.r} > \mathit{orderRound}+1$ and Verify($\mathit{OB}, \mathit{OB.\sigma^O_{set}}$) \textbf{then}
        \State \hspace*{8mm} synchronize missing ordering blocks from $S_{O}$
        \State \hspace*{8mm} update $\mathit{orderRound}$ with the synchronized ordering blocks
        \State \hspace*{4mm} \textbf{elseIf} $\mathit{OB.r} == \mathit{orderRound}+1$ and Verify($\mathit{OB}, \mathit{OB.\sigma^O_{set}}$)
        \Statex{{\footnotesize \color{gray} $\blacktriangleright$ Step 1: handle aggregated results $\mathit{ARGs}$} }
        \State \hspace*{8mm} \textbf{for} $\forall \mathit{ARG} \in \mathit{OB.ARGs}$ \textbf{do}
        \State \hspace*{12mm} \textbf{for} $\forall \mathit{M_{ctx}} \in \mathit{ARG.vote.key()}$ \textbf{do}
        \State \hspace*{16mm} $\Tilde{s} \leftarrow$ get states relevant to $\mathit{ctx}$
        \State \hspace*{16mm} $\mathit{\hat{itxs}}\leftarrow$ get transactions in $\mathit{orderedItxs}$ relevant to $\mathit{ctx}$
        \State \hspace*{16mm} remove $\mathit{\hat{itxs}}$ from $\mathit{orderedItxs}$
        \State \hspace*{16mm} \textbf{if} $\mathit{ARG.vote}[\mathit{M_{ctx}}]$ and $\Tilde{s} \cap \mathit{curAborted} == \emptyset$ \textbf{then}
        \State \hspace*{20mm} remove the relevant states of $\mathit{ctx}$ from $\mathit{curPending}$
        \State \hspace*{20mm} commit $\mathit{ctx}$ and $\mathit{\hat{itxs}}$
        \State \hspace*{16mm} \textbf{else}
        \State \hspace*{20mm} $\mathit{curAborted} \leftarrow$ $\mathit{curAborted} \cup \Tilde{s}$
        \State \hspace*{20mm} abort $\mathit{ctx}$, re-execute and finalize $\mathit{\hat{itxs}}$
        \State \hspace*{12mm} remove states before $\mathit{ARG.round}$ from $\mathit{curAborted}$
        \Statex{{\footnotesize \color{gray} $\blacktriangleright$ Step 2: handle intra-shard transactions} }
        \State \hspace*{8mm} \textbf{for} $\forall \mathit{EB_{dst}} \in \mathit{OB.\hat{EB}_{dsts}}$ and $\mathit{EB_{dst}}$ belongs to $S_P^{\mathit{sid}}$  \textbf{do}
        \State \hspace*{12mm} $\mathit{\hat{itxs}} \leftarrow$ Retrieve($\mathit{EB_{dst}}$)
        \State \hspace*{12mm} \textbf{for} $\forall \mathit{itx} \in \mathit{\hat{itxs}}$ \textbf{do}
        \State \hspace*{16mm} $\Tilde{s}$, $\mathit{succ} \leftarrow$ execute($\mathit{itx}$)
        \State \hspace*{16mm} \textbf{if} $\Tilde{s} \cap \mathit{curPending}$ != $\emptyset$ \textbf{then}
        \State \hspace*{20mm} $\mathit{orderedItxs} \leftarrow \mathit{orderedItxs} \cup \mathit{itx}$
        \State \hspace*{20mm} \textbf{if} $\mathit{succ}$ \textbf{then}
        \State \hspace*{24mm} $\mathit{curPending} \leftarrow \mathit{curPending} \cup \Tilde{s}$
        \State \hspace*{16mm} \textbf{else}
        \State \hspace*{20mm} finalize $\mathit{itx}$
        \Statex{{\footnotesize \color{gray} $\blacktriangleright$ Step 3: handle cross-shard transactions} }
        \State \hspace*{8mm} $\mathit{\hat{ctxs}} \leftarrow \{\}$
        \State \hspace*{8mm} \textbf{for} $\forall \mathit{M_{ctx}} \in \mathit{OB.\hat{M}_{ctxs}}$ and $\mathit{M_{ctx}}$ involves $S_{P}^{\mathit{sid}}$ \textbf{do}
        \State \hspace*{12mm} $\mathit{\hat{ctxs}} \leftarrow \mathit{\hat{ctxs}}$ $\cup$ Retrieve($\mathit{M_{ctx}}$)
        \State \hspace*{8mm} $\mathit{vote} \leftarrow$ voteCtxs($\mathit{\hat{ctxs}}$, $\mathit{curPending}$, $\mathit{curAborted}$)
        \State \hspace*{8mm} send $\langle$\text{\scriptsize EXEVOTE}, $\mathit{vote}, \mathit{OB.round}\rangle$ to its certification module
        \Statex{{\footnotesize \color{gray} $\blacktriangleright$ Step 4: external verifiability } }
        \State \hspace*{8mm} update the ledger $\mathfrak{L}^{\mathit{sid}}$ of $S_P^{\mathit{sid}}$ after handling $\mathit{OB}$
        \State \hspace*{8mm} broadcast $\langle$\text{\scriptsize COMMIT}, $\mathit{\mathfrak{L}^{sid}.root}, \mathit{OB.round}, \mathit{sid} \rangle_{\sigma_j}$ to $S_P^{\mathit{sid}}$
        \Statex {\footnotesize \color{Peach}$\blacktriangleright$ commit a new ledger}
        \State \textbf{upon} receiving a \text{\scriptsize COMMIT} message $\mathit{ct}$ from Node $k$ \textbf{do}
        \State \hspace*{4mm} $\mathcal{M}[\mathit{ct.round}] \leftarrow$ $\mathcal{M}[\mathit{ct.round}] \cup \mathit{ct}$
        \Statex
        \State \textbf{function} voteCtxs($\mathit{\hat{ctxs}}$, $\mathit{pending}$, $\mathit{aborted}$)
        \State \hspace*{4mm} $\mathit{vote}\leftarrow \{\}$ \Comment{{\footnotesize \color{gray} record $\mathit{VOTE}$ results}}
        \State \hspace*{4mm} \textbf{for} $\forall \mathit{ctx}\in \mathit{\hat{ctxs}}$ \textbf{do}
        \State \hspace*{8mm} $\Tilde{s}$, $\mathit{succ}$ $\leftarrow$ execute($\mathit{ctx}$)
        \State \hspace*{8mm} \textbf{if} $\mathit{succ}$ and $\Tilde{s}\cap \mathit{aborted} == \emptyset$ \textbf{then} 
        \State \hspace*{12mm} update $\mathit{pending}$ with $\Tilde{s}$
        \State \hspace*{12mm} $\mathit{vote}[\mathit{M_{ctx}}]=1$
        \State \hspace*{8mm} \textbf{else}
        \State \hspace*{12mm} update $\mathit{aborted}$ with $\Tilde{s}$  
        \State \hspace*{12mm} $\mathit{vote}[\mathit{M_{ctx}}]=0$
        \State \hspace*{4mm} \textbf{return} $\mathit{vote}$
    \end{algorithmic}
\end{algorithm}

\section{Supplementary Details for Arete}
\label{appendix-details}
\balance

\subsection{An Implementation for the \text{\footnotesize EXECUTE} Stage}\label{appendix-detailed-execution-stage}
This section provides a detailed implementation for the \text{\footnotesize EXECUTE} stage to support lock-free execution based on Arete. This is an implementation for the single transaction execution, we emphasize existing effective transaction execution engines (e.g., Blcok-STM~\cite{blockstm}) can be integrated into Arete to support more complex and functional transactions, as transactions are globally ordered and this execution process is independent to either data dissemination task or ordering task.

The pseudocode of nodes performing the \text{\footnotesize EXECUTE} stage is shown in Algorithm~\ref{alg-execution-node-execution}. Specifically, upon receiving an ordering block $\mathit{OB}$ from $S_O$, node $\mathrm{N}_j^{\mathit{sid}}$ in $S_P^{\mathit{sid}}$ traces the latest consensus round $\mathit{orderRound}$ and uses a synchronizer to ensure not miss any ordering blocks it involves (lines 1-4). Arete adopts a \textit{lock-free execution approach} to handle cross-shard transactions, i.e., processing shards optimistically execute cross-shard transactions without locking involved states. Since finalizing a cross-shard transaction involves two consensus rounds (one for ordering, and the other for aggregating $\mathit{VOTE}$ results), transactions that are ordered between these two consensus rounds may access these executed but not finalized states generated by the cross-shard transaction. To this end, nodes in processing shards maintain (i) $\mathit{curPending}$ that records all states generated by all successfully executed but not finalized transactions; (ii) $\mathit{curAborted}$ that records all states generated by all failed cross-shard transactions; (iii) $\mathit{orderedItxs}$ that records all ordered intra-shard transactions that have not been finalized since they access states in $\mathit{curPending}$.

Finalizing transactions in $\mathit{OB}$ consists of three steps (lines 6-32). First, nodes finalize preceding intra-shard and cross-shard transactions relevant to the aggregated results $\mathit{ARGs}$ in $\mathit{OB}$ (lines 6-17). Recall that $\mathit{ARGs}$ aggregate all $\mathit{VOTE}$ results on preceding ordered cross-shard transactions $\mathit{ctxs_{pre}}$ and indicate that $\mathit{ctxs_{pre}}$ are committed or aborted eventually. Once $\mathit{ctxs_{pre}}$ are finalized, their subsequent relevant transactions can be finalized as well. Then, in the second step, nodes execute the ordered intra-shard transactions in $\mathit{OB}$ (lines 18-27). An intra-shard transaction can be finalized in the current consensus round if it does not involve any state in $\mathit{curPending}$. Otherwise, it is put into $\mathit{orderedItxs}$ and will be finalized until its involved states in $\mathit{curPending}$ are finalized. Next, in the third step, nodes execute and vote the execution results for the newly ordered cross-shard transactions in $\mathit{OB}$ (lines 28-32, 37-47). A cross-shard is labeled as \textit{vote-for-commit} (i.e., 1) if it is executed successfully and does not involve any state in $\mathit{curAborted}$. Apart from finalizing transactions, an extra step is required to achieve external verifiability property (lines 33-36). The external verifiability property ensures that external nodes (e.g., client) can verify the correctness of the finalized ledger $\mathit{\mathfrak{L}^{sid}}$~\cite{momose2021multi, das2022spurt}, and is important for the blockchain field, e.g., some light clients in Ethereum require a Merkle proof to determine if a transaction or state is committed into the blockchain. An external node considers a finalized ledger $\mathit{\mathfrak{L}^{sid}}$ valid if its associated committed message $\mathcal{M}$ includes at least $f_S\cdot|S_P^{\mathit{sid}}|+1$ signatures. Depending on the concrete implementation of the sharding protocol, the term \textit{state} discussed above can be various. For instance, when adopting an account transaction model similar to Ethereum, a state could represent an account (for coarse-grained concurrency) or a variable of a smart contract (for fine-grained concurrency). There are many scalable and efficient solutions~\cite{lu13aria,guo2021releasing,dong2023fine} that can obtain the states involved in the \text{\footnotesize EXECUTE} stage but are considered orthogonal to this paper.

\subsection{Protocol Extension}
\label{appendix-extension-details}
We also introduce several protocol extensions that can be flexibly applied to Arete.

\noindent \textbf{Fairly order across shards}. A prior work Haechi~\cite{zhang2023front} explores a new type of front-running attack in blockchain sharding systems, called cross-shard front-running attack. The cross-shard front-running attack allows the attacker to manipulate the execution order of transactions across different shards so that their transactions can be executed before the victim's transactions. The key idea proposed by Haechi to prevent such cross-shard front-running attacks involves globally ordering transactions with a fair ordering policy before execution. Given that Arete similarly adopts a global ordering of transactions before execution, the fair ordering policy introduced by Haechi can be seamlessly integrated into the \text{\footnotesize ORDER} stage of Arete.


\noindent \textbf{Data compression}. 
While we use the metadata of cross-shard transactions (or transaction batches as discussed in \S~\ref{sec-coe-model-optimization}) in the aggregators $\mathit{ARGs}$ in the above algorithms for illustration purposes, the metadata is refined to the bit strings in our implementation. Specifically, since the cross-shard transactions are explicitly ordered and recorded in the field $\mathit{\hat{M}_{ctxs}}$ of $\mathit{OB}$, we can avoid using the repeated metadata in the field $\mathit{ARGs}$. Instead, we use a sequence of binary bits to represent if cross-shard transactions can be committed or not. For instance, assume an ordering block in consensus round $r_1$ has a list of metadata $\mathit{\hat{M}_{ctxs}}=\{\mathit{M_{ctx_1}}, \mathit{M_{ctx_2}}, \mathit{M_{ctx_3}}\}$, we can set $\mathit{ARG}=\{\mathit{round}: r_1, \mathit{vote}: 110\}$ to represent $\mathit{ctx_1}$ and $\mathit{ctx_2}$ are committed while $\mathit{M_{ctx_3}}$ is aborted eventually, and then each processing shard resolves $\mathit{ARG}$ locally. This can incredibly reduce the size of $\mathit{ARGs}$. For example, when using the SHA-256 algorithm to calculate a hash (i.e., a hash is 256 bits long) in metadata, replacing metadata with bit strings can achieve at least $256\times$ data compression. Similarly, for the $\mathit{VOTE}$ results, we can also compress them with bit strings as their corresponding cross-shard transactions have been explicitly ordered in ordering blocks and can be distinguished without using the metadata format. 

\noindent \textbf{Complex cross-shard smart contracts}. In addition to one-shot cross-shard transactions, multi-shot cross-shard transactions may exist in a sharding system because of complex smart contracts that interact with each other and share their states during execution. When these complex smart contracts are deployed into a sharding system, it can lead to state exchanges between shards, introducing more complicated cross-shard coordination. In Arete, since there is a global order for transactions, processing shards can consistently synchronize all involved states and contract codes via off-chain communication so that each of them can execute multi-shot transactions independently. Such an off-chain synchronization solution for multi-shot cross-shard transactions can be integrated into the \text{\footnotesize{EXECUTE}} stage, but we leave this in our future work. In fact, how to efficiently and securely support complex cross-shard smart contracts is still an open problem for all existing sharding protocols. Instead, many works~\cite{rivet, tao2020sharding,krol2021shard} explore an alternative solution to eliminate complex cross-shard contracts, e.g., deploying contracts based on their dependencies by which there is no contract interaction cross shards. Furthermore, some speculation solutions~\cite{rivet, forerunner,hong2023prophet} are proposed to obtain involved states of contracts before executing them, by which shards are not required to exchange states. We emphasize that these solutions are compatible with Arete.

\noindent \textbf{Adaptive creations of execution blocks.} Currently, Arete only assigns one ordering shard to handle transactions from all processing shards (i.e., ordering transactions and coordinating cross-shard commitments). A fundamental concern is that the ordering shard could become the bottleneck with more processing shards occupying the bandwidth of the ordering shard. An effective solution is to let processing shards create larger execution blocks, by which the size of data the ordering shard has to handle becomes smaller because each processing shard creates fewer certificate blocks. In other words, the overheads are amortized by processing shards. This solution is feasible with a reasonable incentive mechanism: nodes in processing shards can get more transaction fees if their execution blocks package more transactions; 2) the ordering shard charges processing shards less if an execution block contains more transactions.

\noindent \textbf{Leaderless cross-shard coordination.} Currently, Arete assigns the ordering shard to coordinate/lead the finalization of cross-shard transactions (\S~\ref{sec-coe-specification}). An alternative is to employ processing shards to communicate with each other directly to finalize cross-shard transactions, formulating a so-called \textit{leaderless coordination}. Specifically, instead of sending $\mathit{VOTE}$ results to the ordering shard, all processing shards need to broadcast their $\mathit{VOTE}$ results to the others. Then each processing shard can locally aggregate $\mathit{VOTE}$ results and finalize cross-shard transactions based on the established global order. This leaderless cross-shard coordination may reduce the confirmation latency of cross-shard transactions as it does not rely on the ordering shard to propose a new block. However, this all-to-all communication brings significantly more communication overhead compared to the current leader-based cross-shard coordination.

\begin{figure}[ht]
    \setlength\abovecaptionskip{-0.0\baselineskip}
    \setlength\belowcaptionskip{-1.0\baselineskip}
    \centering
    \includegraphics[width=3.5in]{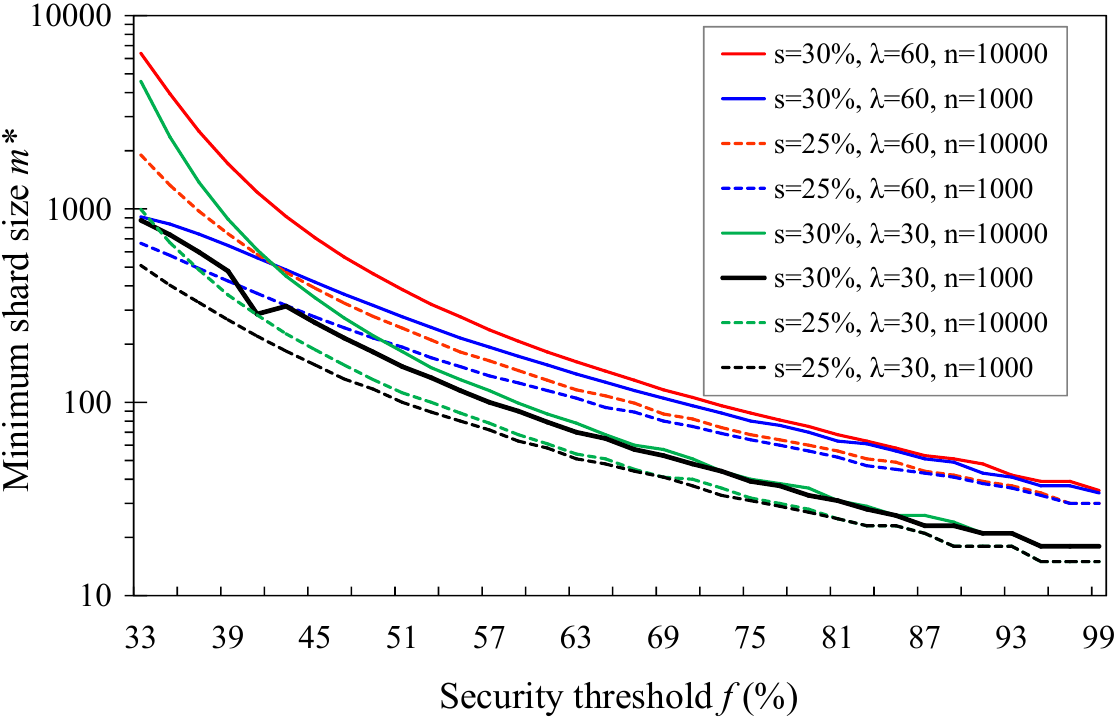}
    \caption{Evaluation for minimum shard size under different settings.}
    \label{fig-relation}
    \vspace{-5mm}
\end{figure}

\section{Minimum Shard Size Evaluation}
\label{appendix-size-evaluation}
Using Equations (\ref{eq-formation}) and (\ref{eq-mini}), we can calculate the minimum shard size $m^*$ for a blockchain sharding system with the given total number of nodes $n$, total fraction of Byzantine nodes $s$, security threshold of a shard $f$, and the security parameter $\lambda$. 
This section demonstrates the impacts of $n$, $s$, $f$, and $\lambda$ on $m^*$, providing insights to interested readers who want to design their own optimal sharding systems. 

Figure~\ref{fig-relation} shows the relationships among $n$, $s$, $f$, $\lambda$, and $m^*$. We can observe that to reduce the minimum shard size $m^*$, one can (i) have fewer nodes in the system (i.e., decreasing $n$), (ii) sacrifice the total fault tolerance of the system (i.e., decreasing $s$), (iii) compromise the security level of forming a secure shard (i.e., assume a smaller $\lambda$), or (iv) enhance the security fault tolerance $f$ (as did in Arete). 

We can further conclude from the observation(i) that when sampling nodes to form shards without replacement (i.e., $n$ decreases after a shard is created from the node pool), the minimum shard size $m^*$ will be reduced for a newly formed shard if $s$ is fixed. Since nodes are randomly sampled to form a shard, the Byzantine ratio in the sampled shard (i.e., the actual ratio of Byzantine nodes in the shard) is close to the overall Byzantine ratio $s$ with high probability~\cite{gearbox}. In other words, $s$ can be kept nearly fixed during the one-by-one creation of shards without node replacement. Therefore, based on the above observation, $m^*$ can be continuously reduced during the shard formation process. However, in this paper, we only consider the maximum $m^*$ that is calculated for the first created shard. As a result, even though we calculate the number of shards $k$ by simply dividing $n$ by $m^*$, the calculated $k$ is the lower bound of the number of shards that a sharding system can actually create. However, since existing sharding protocols all ignore the change of $m^*$ and use the calculated lower bound $k$ to indicate the number of shards, we also use the same calculation in this paper for a fair comparison.

\end{document}